\documentclass[12pt]{article}

\usepackage[letterpaper, margin=0.8in]{geometry}
\usepackage{authblk}
\newcommand{\email}[1]{\texttt{\small #1}}
\usepackage{amsfonts}
\usepackage{amsmath}
\usepackage{natbib}
\usepackage{graphicx} 
\usepackage{enumitem}
\usepackage{setspace}
\usepackage{parskip}
\usepackage{outlines}
\usepackage{color}
\usepackage{bm}
\usepackage{bbm}
\usepackage{float}
\usepackage{xcolor}
\usepackage[raggedright]{titlesec}

\usepackage{cancel}
\usepackage{mathtools}

 \definecolor{bluegray}{rgb}{0.04,0,0.7}
 \definecolor{darkbrown}{rgb}{0.40,0.2,0.05}
 \definecolor{forestgreen}{RGB}{34, 139, 34}
 \definecolor{codegreen}{rgb}{0,0.6,0}
 \definecolor{codegray}{rgb}{0.5,0.5,0.5}
 \definecolor{codepurple}{rgb}{0.58,0,0.82} \definecolor{backcolour}{rgb}{0.95,0.95,0.92}

 \usepackage[colorlinks=true,citecolor=codegreen,linkcolor=darkbrown,urlcolor=blue,breaklinks]{hyperref}
\usepackage{cleveref}

\usepackage{etoolbox}
\usepackage{caption}
\usepackage{subcaption}
\newtheorem{proposition}{Proposition}
\newtheorem{definition}{Definition}
\newtheorem{proof}{Proof}

\DeclareMathOperator{\E}{\mathbb{E}}
\newcommand{\ind}{\perp\!\!\!\!\perp} 
\newcommand{\opdo}{\text{do}} 
\newcommand{\cov}{\text{cov}} 
\newcommand{\cor}{\text{cor}} 
\newcommand{\var}{\text{var}} 
 
\newcommand{\ACE}{\text{ACE}}

\newcommand{\betaIV}{\beta_\text{IV}}

\makeatletter
\def\ourassump#1{\expandafter\@ourassump\csname c@#1\endcsname}
\def\@ourassump#1{\ifcase#1\or\textbf{Surgicality}\or \textbf{Aggregation Restriction}\or\textbf{Value Independence}\fi}
\makeatother
\AddEnumerateCounter{\ourassump}{\@ourassump}{Surgicality}

\makeatletter
\def\ivassump#1{\expandafter\@ivassump\csname c@#1\endcsname}
\def\@ivassump#1{\ifcase#1\or\textbf{Exchangeability}\or \textbf{Relevance}\or\textbf{Exclusion Restriction}\fi}
\makeatother
\AddEnumerateCounter{\ivassump}{\@ivassump}{Exchangeability}

\newcommand{\setword}[2]{%
    \phantomsection #1\def\@currentlabel{\unexpanded{#1}}\label{#2}%
}
\makeatother

\title
{\LARGE \textbf{Lost in Aggregation: \\
The Causal Interpretation of the Instrumental Variable Estimand}} 

\author{
    \textbf{Danielle Tsao}$^1$, \textbf{Krikamol Muandet}$^2$, \textbf{Frederick Eberhardt}$^3$, \textbf{Emilija~Perković}$^1$ \\
    \medskip
    \small $^1$University of Washington, Seattle, USA \\
    \small $^2$CISPA Helmholtz Center for Information Security, Saarbrücken, Germany \\
    \small $^3$California Institute of Technology, Pasadena, USA \\
    \medskip
    \email{dltsao@uw.edu} }

\date{}

\begin{document}

\maketitle

\abstract{Instrumental variable estimation has emerged as a standard approach to mitigating confounding bias in the social sciences and epidemiology, where conducting randomized experiments can be too costly or infeasible. However, justifying the validity of the instrument is frequently challenging. We highlight a problem generally neglected in arguments for instrumental variable validity: the presence of an ``aggregate treatment variable'', where the treatment (e.g., education, GDP, caloric intake) is composed of finer-grained, unobserved components that each may have a different effect on the outcome. While the aggregation problem itself is general, our focus is on instrumental variable estimation in a linear setting, the regime underlying much of applied instrumental variable practice. We show that the causal effect of an aggregate treatment is generally ambiguous, as it depends on how an intervention on the aggregate is instantiated at the component level. We formalize this relation using the aggregate-constrained component intervention distribution (ACID).~We then identify two key conditions under which standard instrumental variable estimators identify the aggregate effect. The contrived nature of these conditions implies major limitations on the interpretation of instrumental variable estimates based on aggregate treatments and highlights the need for a broader justificatory base for the exclusion restriction in such settings.}

\maketitle


\section{Introduction}\label{sec:intro}

Instrumental variables (IV) allow practitioners to mitigate confounding bias when estimating the causal effect of a treatment $A$ on some outcome $Y$ from observational data alone. The technique is widely used across disciplines, notably in the social sciences \citep{angrist1991does,acemoglu2008income}  and epidemiology \citep{rashad2006structural,nordestgaard2012effect}, where randomized experiments can be too costly or even impossible to conduct. 
Famously, \cite{angrist1991does} use the quarter-of-birth 
as an instrument to estimate the effect of education on wages. \cite{acemoglu2008income} use the national savings rate as an instrument to estimate the causal effect of gross domestic product (GDP) on the degree of democracy of a country. \cite{rashad2006structural} uses restaurant meal prices as an instrument for estimating the causal effect of caloric intake on body mass index (BMI). \cite{nordestgaard2012effect}, in turn, use particular genotypes as instruments for estimating the causal effect of BMI on ischemic heart disease.

IV estimation relies on a random variable $I$, called an instrument, that stands in a particular relationship to the treatment and outcome: $I$ is associated with the treatment $A$ (relevance), but not with the outcome $Y$ (exchangeability), except through its relationship with $A$ (exclusion restriction) (\citealp{reiersol1945confluence}). While the relevance condition can be tested explicitly, exclusion and exchangeability are not testable unless one makes additional assumptions about the IV set-up; see, e.g., the approaches by \citet{kitagawa2015test,burauel2023evaluating}. As a result, establishing the validity of the instrument constitutes a substantive epistemic challenge, grounded in expert knowledge and background assumptions, and often highly contested; see, e.g., \citet{bound1995problems} and \citet{buckles2013season} with respect to the analysis in \citet{angrist1991does}.  

In the above examples, the treatment (education, GDP, caloric intake and BMI, respectively) is an aggregate of many, {generally unobserved or at least unmeasured,} finer-grained quantities into a single univariate variable, where the individual components can have markedly different effects on the outcome. For example, as \citet{caetano2025causal} note, \emph{education}, measured in units of time, summarizes many different types of education in different domains, which very plausibly have different influences on future income. Given that the IV estimates are used to inform real-world interventions, two important questions arise:
\begin{enumerate}
    \item[(a)] What is the nature of the causal effect of interest when the treatment is an aggregate?, and 
    \item[(b)] Under what circumstances does the classical IV estimator identify this aggregate causal effect? 
\end{enumerate}

The first question relates closely to fundamental assumptions about the nature of causal effects used in the literature on potential outcomes. Originally articulated as part of the \emph{stable unit treatment value assumption} (SUTVA), \citet{rubin1986comment} requires for the estimation of a causal effect that ``the value of $Y$ for unit $u$
when exposed to treatment $t$ will be the same no matter what mechanism is used to assign treatment $t$ to unit $u$ [...]''. \citet{vanderweele2009concerning} isolates the concern more precisely in what he calls \emph{treatment variation irrelevance}, namely, that the potential outcome takes the same value irrespective of what means are used to instantiate the cause. While VanderWeele considers different protocols of instantiating the treatment (such as different ways of performing medical surgery), the point similarly applies to aggregate treatments where an aggregate treatment value may be instantiated by a variety of different configurations of its components. In many social science settings where IV estimation techniques are used, it is highly plausible that treatment variation irrelevance is violated. 

The second question connects back to the literature in economic theory on aggregation bias, which investigates the mismatch between estimands for macro-relations and aggregated estimands for the corresponding micro-relations in the context of common estimation methods such as 
{ordinary least squares} (OLS) \citep{theil1954linear,lichtenberg1990aggregation} and two-stage least squares (2SLS) \citep{theil1959aggregation}. As this literature highlights, the relation between aggregate and components has to be highly constrained in order to attribute meaningful causal effects to aggregate treatments.

While challenges due to aggregation can arise for causal inference in general, we focus here on the IV setting as it has become the go-to method to infer a causal quantity from purely observational data. We consider specifically the linear case where the aggregate is observed but not the components, in order to capture the regime in which IV is predominantly applied.

We make the following contributions: We formalize the linear IV setting with an aggregate treatment {and unobserved components} in Section~\ref{sec:setting} and derive the standard IV estimand applied to this setting. In Section~\ref{sec:groundtruth} we then motivate and justify what we take to be the ground-truth causal effect of the aggregate treatment on the outcome that would provide a suitable basis for policy advice, addressing the first question above. 
Section~\ref{sec:2sls_aggr_match} then identifies what additional assumptions about the aggregate setting are necessary {and sufficient} in order for the IV estimand to correctly identify the aggregate causal effect, addressing the second question. 
We refer to these assumptions as \emph{proportional aggregation} (Section~\ref{sec:propagg}) and \emph{instrument-tuned interventions} (Section~\ref{sec:instaggeq}). Proportional aggregation amounts roughly to the assumption that either each component variable has the same effect on the outcome, or,  that the effect on the outcome is mediated by the aggregate treatment. That is, proportionality renders the aggregation claim close to trivial. In contrast, instrument-tuned interventions require a highly contrived intervention policy, which is implausible in real-world practical cases.  

In Section~\ref{subsec:simulation} we show how IV estimates diverge from the target aggregate causal effect when the particular assumptions of Sections~\ref{sec:propagg} and \ref{sec:instaggeq} are not satisfied.
In Section~\ref{sec:diagnosis} we connect these findings to the exclusion restriction for IV validity to show that these circumstances are untestable in practice and that, consequently, assumptions about aggregation need to be addressed when the exclusion assumption is justified for a domain of application. We also argue that standard techniques to mitigate violations of the exclusion restriction will in general not apply in this aggregate setting. Finally, in Section~\ref{sec:extensions} we consider several related IV settings {and set out what our results imply for applied practice.}
We conclude with a strong note of caution: If IV estimates that involve aggregate treatment variables are taken to provide sound policy advice, then rather strong assumptions about the nature of the aggregation or the policy intervention have to be justified.

\begin{figure}
  \centering  
\includegraphics[width = 10cm]{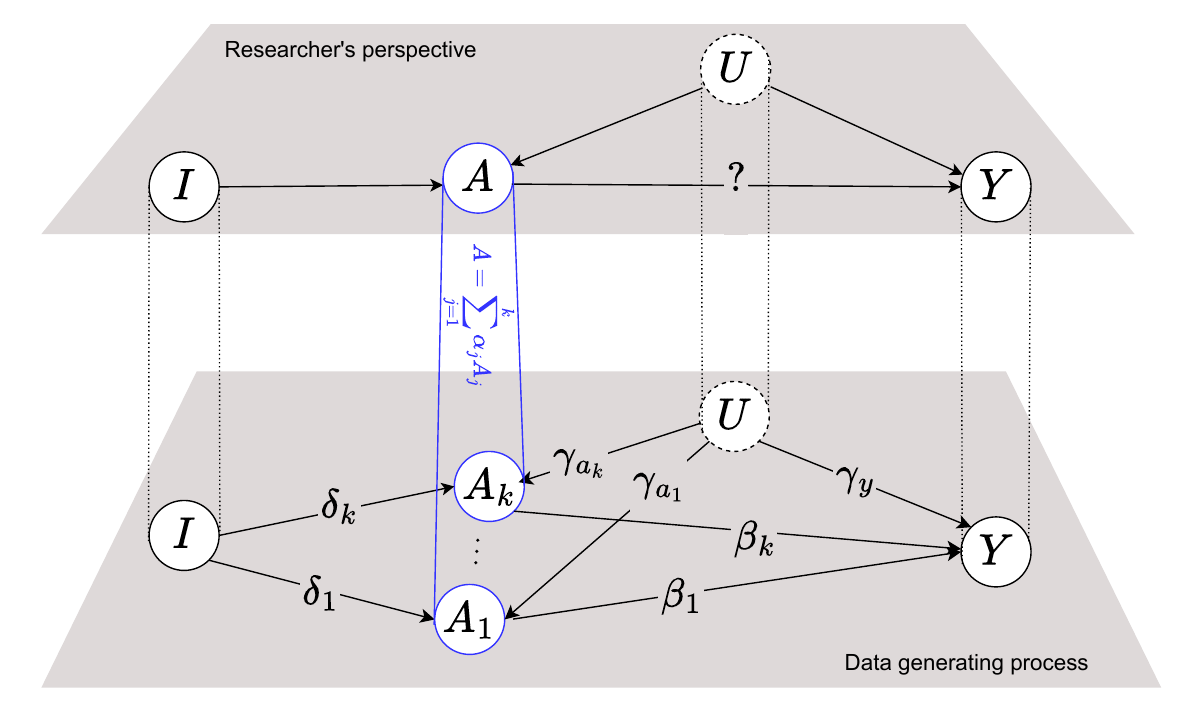}
\caption{The aggregate setting: Comparison of the classic IV setting as posited by the researcher (top) and the true data generating process (bottom). {The researcher does not observe the components $A_1,\ldots, A_k$ or the confounder $U$.}}
 \label{fig:case11}
\end{figure}

\section{The Aggregate Setting}\label{sec:setting}

Consider the setting illustrated in Figure~\ref{fig:case11}: From the researcher's perspective (top) it is a standard IV setting. To estimate the causal effect of treatment $A$ on outcome $Y$ the researcher uses an instrumental variable $I$ to mitigate the potential confounding due to unobserved variable $U$ where $I$ only directly causally affects $A$ and is not confounded with $Y$. However, suppose that unbeknownst to the researcher, treatment $A$ is made up of $k>1$ components $A_1,\ldots, A_k$ that may be subject to different influences $\delta_{j}$ and $\gamma_{a_j}$ from the instrument $I$ and confounder $U$, respectively, and that may have different effects $\beta_j$ on outcome $Y$ (bottom).    That is, for the linear case the data generating process is given by the following structural causal model (SCM):
\begin{subequations}
\label{scm:generalcase} 
\begin{align}
    U &\leftarrow \epsilon_u; \quad I \leftarrow \epsilon_i  \label{eq:gen1}\\
    A_j &\leftarrow \delta_{j} I + \gamma_{a_j}U + \epsilon_{a_j}, \quad j \in \{1,...,k\} \label{eq:gen2} \\ 
    A \-\  &= \-\ \sum_{j=1}^k \alpha_jA_j\label{eq:aggconstr} \\
    Y  &\leftarrow \sum_{j=1}^k \beta_jA_j + \gamma_{y}U + \epsilon_y .
\end{align}
\end{subequations}
{Throughout, we take the instrument $I$, the unobserved components $A_1,\dots,A_k$, the
aggregate treatment $A=\sum_j \alpha_j A_j$, and the outcome $Y$ to be continuous, real-valued
variables.}
Intercepts are omitted for simplicity and without loss of generality. The relation~\eqref{eq:aggconstr} between $A$ and its components $A_j$ is stated as an equation, rather than as a causal effect, indicating that this relation is definitional or constitutive, rather than causal, and consequently cannot be broken by experimental intervention. Following \citet{caetano2025causal}, we refer to \eqref{eq:aggconstr} as the \emph{aggregation rule.} For simplicity we consider the aggregation rule to be a simple weighted linear sum, and that the error terms $\epsilon_u, \epsilon_i, \epsilon_y, \epsilon_{a_j}$ for $j\in\{1,\ldots,k\}$ are mutually independent with mean zero and finite variance.

Given that the aggregate setting is linear with a single instrument, all standard IV estimation techniques, such as two-stage least squares (2SLS) \citep{theil1953repeated,basmann1957generalized} and indirect least squares \citep{durbin1954errors, angrist1996identification}, target the same estimand \citep{burgess2017review}. We denote this general linear IV estimand as $\betaIV$.
Applying 2SLS, for example, to derive $\betaIV$ in the aggregate setting amounts to first regressing $A$ on $I$ to obtain linear predictor $\tilde{A}$ of $A$, and then regressing $Y$ on $\tilde{A}$ to obtain the estimand.
For the aggregate setting this implies:
\begin{align}
    \tilde{A} = \frac{\cov(\sum_{j=1}^k \alpha_jA_j,I)}{\var(I)}I, 
\end{align}
and hence
\begin{align}
    \betaIV = \frac{\cov(Y,\tilde{A})}{\var(\tilde{A})} =
    \frac{\cov(Y,I)}{\sum_{j=1}^k \alpha_j\cov(A_j,I)} 
    = \frac{\sum_{j=1}^k \beta_j\delta_j}{\sum_{j=1}^k \alpha_j\delta_j}. \label{eq:2sls}
\end{align}

For a standard IV model as in the top of Figure~\ref{fig:case11}, $\betaIV$ 
is interpreted as the causal effect of the treatment on the outcome (defined formally in Section~\ref{sec:groundtruth}).
However, for the aggregate setting in Figure~\ref{fig:case11}, it is unclear which interventions (if any) correspond to this estimand, since
the intervention on $A$ has to be realized in its components with possibly many ways of instantiating a single value.

\section{The Causal Effect of an Aggregate Treatment}\label{sec:groundtruth}

In linear models, the target causal effect is usually defined as the expected change in outcome $Y$ resulting from a unit change of the intervention on the treatment $A$:
 \begin{align}
  \E[Y \,|\, \opdo(A = a+1)] - \E[Y \,|\, \opdo(A = a)],    \label{eq:ATE-old}
 \end{align}
 {where we use the do-notation $\opdo(A = a)$ of \cite{pearl2009causality} to denote the intervention that sets the value of treatment $A$ to $a$.}
 This type of causal effect is often referred to as the average causal effect, or the average treatment effect \citep{hernancausal}. 

Given that $A$ is an aggregate, an intervention $\opdo(A=a)$ can be instantiated by many combinations of values for the components $A_1,\ldots, A_k$,  with each satisfying the aggregation rule~\eqref{eq:aggconstr}. The same applies to $\opdo(A=a+1)$, which implies that a unit change in $A$ may involve more (or less) than a unit change in any one $A_j$.
  \cite{spirtes2004causal} refer to this type of causal effect as \emph{ambiguous}. The effect is not well-defined as it is dependent on the particular values of the component $A_j$'s that instantiate the intervention $\opdo(A=a)$, or in other words, it violates treatment variation irrelevance \citep{vanderweele2009concerning}. 

In addition to this value-ambiguity of the components, one has to determine whether under the intervention $\opdo(A=a)$ the values of the components $A_j$ are fixed by interventions $\opdo(A_j)$ that make the $A_j$ independent of their causal parents (here represented by $I$ and $U$), or whether $\opdo(A=a)$ merely enforces the aggregation rule on the values of the $A_j$ without making the $A_j$ independent of their causal parents. Both of these considerations highlight the need for an explicit distribution relating the intervention on the aggregate $A$ to the change of its components $A_1,\ldots, A_k$. 
In the following, we give \eqref{eq:ATE-old} a precise meaning for the setting when treatment $A$ is an aggregate.

\subsection{Aggregate-constrained Component Intervention Distribution (ACID) and the Aggregate Causal Effect} \label{sec:acd}

For a given intervention $\opdo(A=a)$ on the aggregate $A$, we introduce the \emph{aggregate-constrained component intervention distribution (ACID)} to describe how the intervention is instantiated in terms of the components $A_1,\ldots, A_k$, and how the components' relations to all other variables in the causal system change under intervention. 

\begin{definition}[Aggregate-constrained component intervention distribution; ACID]\label{def:acid}
For an intervention $\opdo(A=a)$ on the aggregate $A$ in SCM~(1), and $\mathbf{V} = \{I,U,A_1,\dots,A_k,Y\}$, the ACID is given by an interventional distribution:
\begin{align}\label{eq:iidist}
    P(\mathbf{v})_{\opdo(A=a)}  = P(i,u,a_1, ...,  a_k, y \ | \ \opdo(A = a)). 
\end{align}
\end{definition}
We emphasize that \emph{any} causal claim about \emph{any} aggregate treatment presupposes an ACID, even if it is generally left implicit. 

A researcher who is unaware of the aggregated nature of the treatment (recall Figure~\ref{fig:case11}) standardly makes three (implicit) assumptions about the causal effect in \eqref{eq:ATE-old}: that the intervention on $A$ makes $A$ independent of its causes, that $A=a$ can be interpreted in the same way under intervention and observation, and that it does not depend on the specific value $a$. In the aggregate setting these assumptions impose constraints on the ACID:

\begin{enumerate}[label=(\alph*)]
    \item \label{a:surgicality} \textbf{Surgicality} -  
    The intervention $\opdo(A=a)$ is instantiated in terms of (possibly correlated) interventions on the $A_j$ that make the $A_j$ independent of their observational causes ($I$ and $U$). 
    \item\label{a:agg-constr}  \textbf{Aggregation Rule}
    - $\opdo(A=a)$
    respects the aggregation rule, i.e.\ the ACID lacks support for any combination of values of $a_1,\ldots, a_k$ that does not satisfy the aggregation rule~\eqref{eq:aggconstr}.
    \item\label{a:value-indep} \textbf{Value Independence}
    - 
    For all $a, a'$ values that can be assigned:
    \begin{align*}
        \E[Y | \opdo(A = a+1)] - \E[Y | \opdo(A = a)] = \E[Y | \opdo(A = a'+1)] - \E[Y | \opdo(A = a')].
    \end{align*}
    That is, the causal effect of treatment $A$ on $Y$ is independent of the particular value $a$ that $A$ is set to in the intervention $\opdo(A=a)$.
\end{enumerate}
These conditions are not strictly minimal, but any weakening would change the interpretation of the linear causal effect from how it is usually understood. We discuss this in more detail in Section~\ref{sec:assumptions}.
We refer to an ACID that satisfies these conditions as a \emph{valid} ACID, since it ensures the usual interpretation of the linear causal effect:

\begin{definition}[Valid ACID]\label{def:validacid}
An ACID (Definition~\ref{def:acid}) is \emph{valid} if it satisfies all three conditions above: \ref{a:surgicality} surgicality,  \ref{a:agg-constr} the aggregation rule, and \ref{a:value-indep} value independence.
\end{definition}
A valid ACID in \eqref{eq:iidist} will factor in the following way: 
\begin{subequations}\label{eq:acdfactorization}
    \begin{align}
    P(\mathbf{v})_{\opdo(A=a)}  &=P(i, u | \opdo(a))P(a_1, ..., a_k | i, u, \opdo(a)) P(y |a_1, ..., a_k, i, u, \opdo(a)) \label{eq:chain-rule}\\
    &=P(i,u)P(a_1, ..., a_k | \opdo(A=a)) P(y\ |a_1, ..., a_k, u) \label{eq:partial-id}\\
    &= P(i)P(u)P^*(a_1, ..., a_k; a) P(y\ |a_1, ..., a_k, u),
    \label{eq:pstardef}
\end{align}
\end{subequations}
 where we use $
 \opdo(a)$ instead of $\opdo(A=a)$ for brevity, and the chain rule and the surgicality assumption to transform the interventional distributions into their observational counterparts. 
     Note that the intervention value $a$ is a parameter of $P^*$ because the aggregation rule must be satisfied for $a$. 
     Since the $P^*$-term marks the crucial (and only) difference between the ACID and the observational distribution of $\mathbf{V}$, we will refer to both \eqref{eq:iidist} and the specific $P^*$ distribution as the ACID, as long as it is clear from context. 

The SCM corresponding to an intervention on the aggregate $A$ for the system described by \eqref{scm:generalcase} is then given by:
\begin{subequations}
    \begin{align}
    U &\leftarrow \epsilon_u; \quad I \leftarrow \epsilon_i \\
    (A_1, ..., A_k) & \sim P^{*}(a_1, ..., a_k ; a )  
    \label{eq:iidsurg}\\
    Y  &\leftarrow \sum_{j=1}^k \beta_j A_j + \gamma_{y}U + \epsilon_y.
\end{align} 
\end{subequations}

Furthermore, $\E[Y\,|\,\opdo(A = a)]$ is now computed as
\begin{align}
    \E[Y \,|\, \opdo(A=a)] 
    &= 
    {\int} (\beta_1 a_1 + ... + \beta_k a_k) dP^*(a_1, ..., a_k ;a) \\ 
    &= \sum_{j=1}^k \int \beta_j a_j dP^*(a_j ; a), \label{def:aggcausaleff2} 
\end{align}
where $U$ and $\epsilon_y$ can be dropped from the calculation.
 Then for any $P^*$ that in addition satisfies value independence, we arrive at a well-defined \emph{aggregate causal effect} as follows.
 
 \begin{definition}[Aggregate Causal Effect] \label{def:ace} The causal effect of an aggregate treatment $A$ on an outcome $Y$, referred to as the \emph{aggregate causal effect} and denoted by $\ACE(A,Y)$, is taken to be:
 \begin{align}\label{eq:ATE}
     \ACE(A,Y) &= \E[Y \,|\, \opdo(A = a+1)] - \E[Y \,\,| \opdo(A = a)],  
 \end{align}    
where the expectations of $Y$ are computed as in \eqref{def:aggcausaleff2} under a valid ACID. 
 \end{definition}
For the remainder of this paper we take such an aggregate causal effect to be the target of IV estimation in the aggregate setting, since it supports the usual interpretation of a linear causal effect. In particular, for a non-aggregate treatment $A$, the aggregate causal effect corresponds to the standard average causal effect \citep{hernancausal}. 

With the aggregate causal effect defined, we can now formulate our second question of the introduction precisely: \emph{Under what conditions does an investigator facing the aggregate setting in Figure~\ref{fig:case11}, where the $A_j$ are unobserved, correctly identify the {aggregate causal effect} in \eqref{eq:ATE} using classic IV estimation applied to the aggregate $A$?}  More precisely, what are the constraints on SCM~\eqref{scm:generalcase} and its ACID in \eqref{eq:iidsurg} such that the aggregate causal effect corresponds to the IV estimand $\betaIV$ in \eqref{eq:2sls}? 

\subsection{A Gaussian ACID Example}\label{subsec:gaussian_example}
Consider an ACID $P^*$ that, given intervention $\opdo(A=a)$ on the aggregate $A$, sets the values of the components $A_1, ..., A_k$ according to a multivariate
Gaussian distribution:
\begin{align}\label{condnorm:distn}
    \begin{bmatrix}
        A_1 \\ \vdots \\ A_k
    \end{bmatrix}
    \sim N_k \left(
    \begin{bmatrix}
        c_1 + a d_1 
        \\ \vdots \\ 
        c_k + a d_k
    \end{bmatrix}\hspace{-1mm}, 
    \mathit{\Sigma} 
    \right),
\end{align}
where to satisfy the aggregation rule
the constants $c_j, d_j \in \mathbb{R}$, $j \in \{1, \dots, k\}$ need to satisfy
\begin{align}\label{condnorm:mean}
\sum_{j=1}^k \alpha_jc_j = 0, \quad \enspace 
\sum_{j=1}^k \alpha_jd_j = 1,
\end{align}
and the covariance matrix satisfies the constraint given by $\mathbf{\alpha} := (\alpha_1, \dots, \alpha_k)^\top$:
\begin{align}\label{condnorm:sigma}
 \mathbf{\alpha}^\top \mathit{\Sigma} = 0, \text{ where of course $\mathit{\Sigma} \succeq 0$.}
\end{align}

The surgicality of this ACID is satisfied by construction when the factorizations in \eqref{eq:acdfactorization} are satisfied, and 
value independence is satisfied 
as seen by the fact that
\begin{align}
    \eqref{def:aggcausaleff2} =
    \sum_{j=1}^k \int \beta_j a_j dP^*(a_j ;a) 
    = \sum_{j=1}^k \beta_jc_j + a\sum_{j=1}^k\beta_jd_j
\end{align}
and hence by \eqref{eq:ATE}, 
\begin{align}\label{condnorm:ate}
         \ACE(A,Y) =  \sum_{j=1}^k \beta_jd_j, 
\end{align}
which no longer depends on $a$.

In Supplementary Section~\ref{apx:naturalACD} we show how the distribution in \eqref{condnorm:distn} can also, if desired, be parametrized to match the first and second moments of the 
\emph{observational} distribution over $(A_1, ..., A_k)^{\top}$ conditional on the aggregate $A=a$. 
Note from \eqref{condnorm:ate} that the ACE does not depend on the choice of $\mathit{\Sigma}$ beyond \eqref{condnorm:sigma} and thus also holds for the deterministic case where $\mathit{\Sigma} = 0$. 

\subsection{Discussion of the ACID Assumptions} \label{sec:assumptions}

While the assumption of the surgicality of an intervention  
 is standard \citep{pearl2009causality}, it is a strong assumption in the current context: It requires that an intervention on the aggregate amounts to a surgical intervention on each component even though the experimenter may only have direct control over the aggregate. However, any weakening of this condition 
 risks mistaking confounding for a causal effect. To take one extreme, consider the case where the intervention $\opdo(A=a)$ amounts to conditioning on the aggregation rule, i.e, 
\begin{align}\label{eq:nocond}
     P(\mathbf{v})_{\opdo(A=a)} 
    =  P(i,u,a_1, ..., a_k, y \ | \ \sum_{j=1}^k \alpha_ja_j = a). 
\end{align}
Then setting $A=a$ interventionally would amount to waiting for the observational distribution to bring about suitable values of $A_j$ such that the aggregation rule is satisfied. But that mistakes a sampling technique for an experimental intervention. Moreover, if the $A_j$ had no causal effect on $Y$ (i.e., $\beta_j=0;$ $j \in \{1,...,k\}$), but there was confounding between the $A_j$ and $Y$, then such an ACID would incorrectly suggest that there is an aggregate causal effect of $A$ on $Y$ when there is no directed causal path from $A$ or its components to $Y$. 
Note also that, while not relevant in our setting given the full support of the linear Gaussian distribution, in general for an ACID that merely tracks the observational distribution as in \eqref{eq:nocond}, it may be possible that the observational distribution has no support for any value combination of the components $A_1, \ldots, A_k$ that satisfies the aggregation rule for the intervened value $a$, making such out-of-distribution
interventions impossible. But out-of-distribution interventions  are often the motivation for causal estimation in the first place. 

We acknowledge that there may be circumstances where an intervention can be appropriately defined that does not \textit{fully} sever the relationships of the $A_j$ with $U$ and $I$, but we note that many of the above concerns can be generalized if confounding influences are not broken by the intervention.

Satisfying the aggregation rule ensures that the aggregate variable being considered under intervention is in fact the same as that measured under observation. Without this requirement the aggregate causal effect would be highly underdetermined, since the observational relation no longer constrains the interventional one. There may be specific circumstances where modelling a change in the constitutive relations between the $A_j$ and $A$ is appropriate (for example, when a manager wants to increase overall sales and instantiates that by changing the composition of the underlying product palette). In that case the intervention includes a change in the aggregation weights $\alpha_j$. But this example already indicates that under such circumstances, any inference from previous observational data of the aggregate is likely to provide only very limited guidance about the effect of the intervention unless the disaggregated product categories were already tracked in the first place (in contrast to our setting where the $A_j$ are unobserved). Given our focus on IV estimation in observational data, we leave such examples aside here. We note that we assume that the aggregation rule is satisfied for each interventional data point, not merely in expectation of the interventional distribution, {though none of our results hinge on this assumption}.

Finally, we focus on causal effects that satisfy value independence of the particular intervention value because we are interested in the possible interpretations of classic linear causal model assumptions 
where the estimated causal effect is reported in terms of a single parameter that specifies the causal effect independently of the intervention value. 
Unsurprisingly, this demand for value independence of the causal effect is a particularly strong assumption and imposes strong limitations on the ACIDs that can be considered.

\subsection{Related Work}\label{subsec:relatedwork}
\citet{vanderweele2013causal}
also consider variations in how a treatment is instantiated and offer two settings. In their Section~4 they define their causal effect using a known pre-specified distribution over the different versions of the treatment. This distribution is what we refer to as the ACID. The causal effect is then a mixture of the effects of the components according to this distribution. In their Section~5, which more explicitly considers aggregate treatments, they consider as their ACID the observational distribution over the components conditional on the aggregation rule being satisfied. A similar approach is taken by \citet{zhu2024meaningful}. As we discussed in the context of such an ACID in \eqref{eq:nocond}, this approach requires some care in ensuring that the interventional value is within the observational support, and that the intervention on the components still breaks confounding. In Supplementary Section~\ref{apx:naturalACD} we consider such an ACID explicitly. Neither of these articles consider the conditions under which such causal effects can be accurately estimated using IV techniques. 

{
Closest to our work is \citet{harris2022interpreting}, who considers the IV setting and studies the returns to education through a factor-model approach that interprets IV estimates under violations of SUTVA. The most immediate difference is the support of the variables: we consider a continuous aggregate treatment $A$ with continuous components $A_1,\dots,A_k$, whereas \citet{harris2022interpreting} considers a binary treatment with binary, \emph{mutually exclusive} component treatments. This mutual-exclusivity assumption leads to a different perspective on aggregation: the SUTVA violation is 
an artifact of the instrument, but unlike in our setting, would not be a concern under an intervention.
A separate ``ground-truth'' local average treatment effect is then defined for each component. We instead allow several components to be active whether or not an instrument is present, so that treatment variation irrelevance \citep{vanderweele2009concerning} can fail even under randomization. A further difference in emphasis is that our target of interest is the aggregate causal effect itself, because this is what past studies have reported, whereas \citet{harris2022interpreting} targets the component-level local effects.} 

\citet{caetano2025causal} also consider the causal interpretation problem of aggregate treatments, but do not provide a definition for interventions on an aggregate variable. 
Instead they propose various types of averages over the effects of the components as alternative estimands. Their ``aggregate average treatment effect on
the treated'' (AATT), for example, can be understood as the $\ACE(A,Y)$ under the natural ACID. This is a complementary result to the current work, as \cite{caetano2025causal} arrive at the AATT without working backwards from the implicit assumptions made by researchers in aggregate settings, as we do here. 

In the aggregate setting we are considering, the $\betaIV$ estimand derived in \eqref{eq:2sls} depends on $\delta$'s which are the causal effects of the instrument on the treatment components. Consequently, distinct instruments are likely to imply different $\betaIV$ estimands that may or may not correspond to the $\ACE(A,Y)$, see \eqref{condnorm:ate}. This mirrors the heterogeneous treatment effects literature, where different instruments act on different subpopulations and hence identify different local average treatment effects (LATE, \citealp{imbens1994identification}). But the same heterogeneity can also arise due to aggregation within each unit, rather than resulting from different subpopulations across units. \cite{iong2024latent}  and \citet{caetano2025causal} make this connection explicitly; \cite{iong2024latent} through their investigation of Mendelian randomization studies and the distinct mechanistic pathways by which an instrument may influence the outcome, and \citet{caetano2025causal}
through the following note: ``when treatments are aggregated, what appears to be treatment effect heterogeneity may instead reflect  [...] heterogeneity arising from distinct underlying components of the treatment itself. [...] Crucially, this heterogeneity is not across individuals, but within the same individual under different sub-treatments''. 

Like \cite{caetano2025causal}, we will focus on modelling aggregate treatments per unit rather than explicitly modelling heterogeneity between units; however we argue that our framework offers a structural mechanism for the latter as well. Indeed, one may consider clustering a treated population into subpopulations such that each group takes on a valid ACID under intervention. If different ACIDs are instantiated in different groups, this results in a mixture of ACIDs across the population, yielding distinct local effects in each subpopulation.

The approach in \citet{beckers2020approximate} resembles ours here in that they explicitly denote the ACID describing the relation between intervened aggregate and its components as their \emph{intervention distribution}, but they do not relate their formal framework to effect estimation.

\section{When Do IVs Correctly Identify the Aggregate Causal Effect?}\label{sec:2sls_aggr_match}
We next consider settings where the IV estimand may correctly identify the causal effect. The first of these, which we refer to as \emph{proportional aggregation}, does not depend on the specific form of the ACID; any valid ACID will do. We explore this setting in Section \ref{sec:propagg} below.

Next, using the Gaussian ACID from Section \ref{subsec:gaussian_example}, we show that there are specific ACIDs that ensure that the IV estimand corresponds to the aggregate causal effect. We refer to these settings 
as \emph{instrument-tuned interventions} and explore them in Section \ref{sec:instaggeq}.

\subsection{Proportional Aggregation}\label{sec:propagg}

 To motivate the case of {proportional aggregation}, consider the aggregate setting with $\beta_j = \tau$ and $\alpha_j=1$ for all $j$. In this case, the components $A_j$ simply split the aggregate $A$ into parts that have the same causal effect $\tau$ on $Y$ and so it does not matter how an intervention $\opdo(A=a)$ is instantiated 
in terms of its components. A generalization of this case is what we refer to as \emph{proportional aggregation}. 

{
\begin{definition}[Proportional Aggregation]\label{def:propagg}
The aggregate setting in SCM~\eqref{scm:generalcase} satisfies \emph{proportional aggregation} if the component--outcome effects are proportional to the aggregation weights; that is, there exists $\tau\in\mathbb{R}$ such that
\[
   \beta_j = \tau\,\alpha_j \quad\text{for all } j\in\{1,\dots,k\},
\]
equivalently $\beta_1/\alpha_1 = \cdots = \beta_k/\alpha_k = \tau$.
\end{definition}}

In this case the IV estimand in \eqref{eq:2sls} is
\begin{align*}
    \betaIV =
    \frac{\sum_{j=1}^k\tau\alpha_j\delta_j}{\sum_{j=1}^k \alpha_j\delta_j} = \tau,
\end{align*}
and the aggregate causal effect under proportional aggregation is also
\begin{subequations}
\begin{align}
     \ACE(A,Y) 
     &= \E\left[\tau\sum_j \alpha_j A_j  \bigg|  \opdo\big(A=a+1\big) \right] - 
     \E\left[\tau\sum_j\alpha_j A_j  \bigg|  \opdo\big(A=a\big) \right] \\
     &= \tau(a+1) - \tau a = \tau. 
\end{align}
\end{subequations}

Notice that in this setting, as long as the aggregation rule is respected, any valid ACID can be assumed to relate the intervention on the aggregate $A$ to its components $A_1, \ldots, A_k$. \

There are several ways to interpret the assumption of proportional aggregation. In the simplest case where $\alpha_j=c$ for some constant $c$, it just means that all components of $A$ have the same causal effect on $Y$. More generally, proportional aggregation amounts to an assumption that the components have the same causal effect once rescaled by their share of the aggregate. Alternatively, since under proportional aggregation $A$ (together with $U$) screens off the $A_j$ from $Y$ (i.e.\ $A_j \ind Y | A, U$), the aggregate $A$ can also be interpreted as mediating all the causal effects from $A_j$ to $Y$. So, instead of $Y$ being an effect of the $A_j$ (and the confounder) in SCM~\eqref{scm:generalcase}, we would have:
\begin{align}
  Y  &\leftarrow {\tau}A + \gamma_{y}U + \epsilon_y. 
\end{align}
This is a common way of interpreting aggregate treatments: the effect of the components is mediated by the aggregate. \cite{gabriel2024impact} take this route, as do the approaches that develop ``Bartik instruments'' (see e.g.,\ \citealt{goldsmith2020bartik}). In all these cases there only is a causal effect between the aggregate and the outcome, but no component causal effects as the $\beta_j$ in our setting.

Thus, the assumption of proportional aggregation offers an interpretation of the aggregate setting in  Figure~\ref{fig:case11} that would ensure that the IV estimate can support a policy intervention no matter how the intervention on the aggregate is instantiated in the components. Using our example from the introduction, proportional aggregation would imply that while \emph{education} is made up of many different types of learning, for the assessment of the effect on future wages, one can interpret the IV analysis as assuming that the overall duration of education is the relevant cause, while the specific types of education are irrelevant. Formulated in this way it becomes obvious that this is a rather strong assumption about the nature of the causal effects of different types of education that a student experiences in years K-12: the claim would be that it doesn't matter to future wages whether one adds extra time in kindergarten or in high school, both would be equally beneficial.

\subsection{Instrument-tuned Interventions}\label{sec:instaggeq}

While the assumption of proportional aggregation makes the ACID irrelevant, the second set of scenarios  we raise here depends on particular ACIDs that guarantee that the classic IV estimand corresponds to the aggregate causal effect. We refer to these scenarios as \emph{instrument-tuned interventions} for reasons that will become apparent. 

We focus on the  Gaussian ACID introduced in Section~\ref{subsec:gaussian_example}.
From previous computations, we have that the IV estimand in \eqref{eq:2sls} will match the aggregate causal effect in \eqref{condnorm:ate} when 
\begin{align}\label{iti:equality}
\betaIV = \frac{\sum_{j=1}^k \beta_j\delta_j}{\sum_{j=1}^k \alpha_j\delta_j} \quad = \quad \sum_{j=1}^k \beta_jd_j = \ACE(A,Y). 
\end{align}
Consider the form of the Gaussian ACID in \eqref{condnorm:distn}. Equation \eqref{iti:equality} adds a linear constraint to parameters $d_1, \dots, d_k$, in addition to the constraint in \eqref{condnorm:mean}. The only constraints on the parameters $c_1, \dots, c_k$ and the covariance matrix $\mathit{\Sigma}$ are \eqref{condnorm:mean} and \eqref{condnorm:sigma} in Section~\ref{subsec:gaussian_example}.

As a result there are many possible choices for the parameter values. Only one of these ensures the equality in \eqref{iti:equality} for every value of $\beta_1, \dots, \beta_k$, namely that in which, for all $j \in \{1, \dots, k\}$,
\begin{align}\label{eq:instr_surg_eq}
 d_j = \frac{\delta_j}{\sum_{l=1}^k \alpha_l\delta_l}, 
\end{align}
with $c_1, \dots c_k$ and $\mathit{\Sigma}$ chosen according to Section~\ref{subsec:gaussian_example}.
In words, the above solution implies that the IV estimand identifies the true aggregate causal effect when the slope of the expectation vector of the ACID is equal to the (normalized) causal effect vector of the instrument with respect to the $A_j$'s. 

{
\begin{definition}[Instrument-tuned Intervention]\label{def:iti}
A valid Gaussian ACID~\eqref{condnorm:distn} with mean-slope parameters $d_1,\dots,d_k$ is an \emph{instrument-tuned intervention} if
\[
   d_j = \frac{\delta_j}{\sum_{l=1}^k \alpha_l \delta_l} \quad\text{for all } j\in\{1,\dots,k\};
\]
that is, the slope of each component's interventional mean equals the (normalized) effect of the instrument $I$ on that component. 
\end{definition}}

The \emph{instrument-tuned intervention} and all other solutions to the parameter constraints 
are highly contrived in the sense that they require the surgical intervention that characterizes the aggregate causal effect of $A$ on $Y$ to align with the instrument-treatment relationship. It would be a matter of extraordinary coincidence, {or would require detailed knowledge of the instrument-component relation,} to  achieve such an ACID in practice.

 When the aggregate consists of more than two components such that some but not all  $j \in \{1, \dots, k\}$ have a proportional relationship $\beta_j/\alpha_j = \tau$ for some $\tau$, one might also consider a \textit{partially instrument-tuned intervention}, which only instrument-tunes the slopes of those components $A_j$  for which $\beta_j/\alpha_j \neq \tau$.
For instance, suppose that for $j = \{1, \dots, l \}$, $1 < l < k$, we have
 ${\beta_j}/{\alpha_j} = \tau$. Then, $\betaIV = \ACE(A,Y)$ for partially instrument-tuned ACIDs such as a Gaussian ACID where \eqref{eq:instr_surg_eq} holds for $d_j$ such that $j \in \{l+1, \dots, k\}$, while the remaining parameters $d_1, \dots, d_l$ and $c_1, \dots, c_k$ and $\mathit{\Sigma}$ are such that \eqref{condnorm:mean} and \eqref{condnorm:sigma} are satisfied. More details can be found in Supplementary Section \ref{app:partial-iti}.

\begin{definition}[Partially Instrument-tuned Intervention]\label{def:piti}
Fix $\tau\in\mathbb{R}$ and write $\beta_j = \tau\alpha_j + r_j$. A valid Gaussian ACID is a \emph{partially instrument-tuned intervention} if, for every $j\in\{1,\dots,k\}$, either $r_j = 0$ or $d_j = \delta_j/\sum_{l=1}^k\alpha_l\delta_l$. \end{definition}
Proportional aggregation and (fully) instrument-tuned interventions are the extreme cases in which the first or the second condition holds for all $j$.

We can now state in general when the IV estimand recovers the aggregate causal effect. The result holds for any valid ACID whose component means are affine in the intervention value, i.e. $\E[A_j\mid\opdo(A=a)] = c_j + a\,d_j$ where $c_j, d_j \in \mathbb{R}$ (as for the Gaussian ACID of Section~\ref{subsec:gaussian_example}). 

\begin{proposition}[IV estimand recovers the aggregate causal effect]\label{prop:char}
Consider SCM~\eqref{scm:generalcase} and a valid ACID with $\E[A_j\mid\opdo(A=a)] = c_j + a\,d_j$, so that $\ACE(A,Y) = \sum_{j=1}^k \beta_j d_j$. Let $s_j = \delta_j / \sum_{l=1}^k \alpha_l\delta_l$, and for any $\tau\in\mathbb{R}$ write $r_j = \beta_j - \tau\alpha_j$. Then
\[
   \ACE(A,Y) - \betaIV \;=\; \sum_{j=1}^k r_j\,(d_j - s_j),
\]
and hence $\betaIV = \ACE(A,Y)$ if and only if $\sum_{j=1}^k r_j(d_j - s_j) = 0$.
\end{proposition}

Proposition~\ref{prop:char} shows that $\betaIV = \ACE(A,Y)$ exactly when $\sum_j r_j(d_j - s_j) = 0$. {The proof is provided in Supplementary Section~\ref{apx:aggexclproof}.}  We have discussed the more easily interpretable cases in which this equality occurs above: for \emph{proportional aggregation} (Definition~\ref{def:propagg}) we have $r_j = 0$ for all $j$ and the equality holds for \emph{every} valid ACID; for an \emph{instrument-tuned intervention} (Definition~\ref{def:iti}) we have $d_j = s_j$ for all $j$ and the equality holds for that \emph{particular} ACID; and a \emph{partially instrument-tuned intervention} (Definition~\ref{def:piti}) combines the two. 

\subsection{Illustrations by Simulation}\label{subsec:simulation}

\begin{figure}
\centering
 \includegraphics[width = \linewidth]{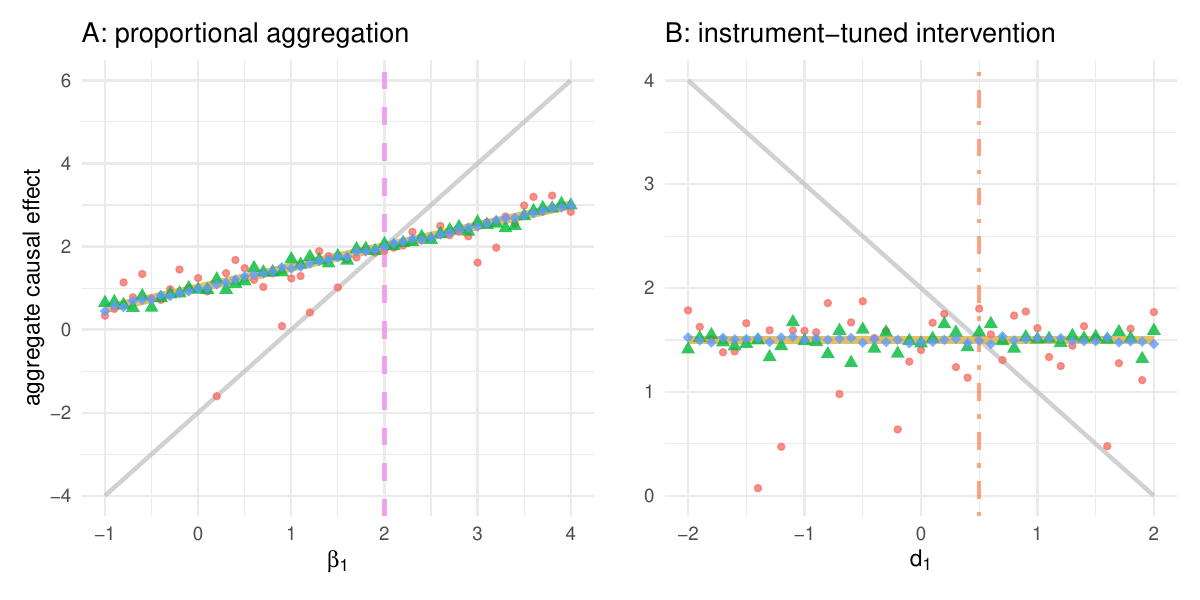}
    \caption{A: IV estimates 
    for different sample sizes $n$ and edge weights $\beta_1$. The vertical purple dashed line indicates 
    proportional aggregation. 
    B: IV estimates for 
    different sample sizes $n$ and ACID slopes $d_1$. The vertical orange dash-dotted line denotes 
    the instrument-tuned intervention. In both plots, $\betaIV$ is denoted by the yellow line and the $\ACE(A,Y)$ 
    is denoted by the grey line.  
   The IV estimates are denoted by: red circles ($n=10$), green triangles ($n =100$), blue diamonds ($n=1000$). 
   }
         \label{fig:sim_mismatch}
\end{figure}

We conclude this section by illustrating the mismatch that can occur between the aggregate causal effect \eqref{condnorm:ate} and the IV estimand \eqref{eq:2sls} as the model coefficients diverge from one of the two settings described above: (a) proportional aggregation and (b) instrument-tuned intervention.
In both cases, we simulate data from SCM~\eqref{scm:generalcase} for $k =2$ components. The coefficients $\alpha_1, \alpha_2, \gamma_{a_1}, \gamma_{a_2}, \gamma_y,$ and $\delta_1$, $\delta_2$ are all set to 1 in the below simulations, the coefficient $\beta_2$ is set to $2$, and all error variances are set to $1$. We generate samples of three different sizes $n$; $n \in \{10, 100, 1000\}$. For simplicity, to compute the aggregate causal effect we choose a Gaussian ACID with $c_1 = c_2 = 0$ in \eqref{condnorm:mean}. The remaining parameters are chosen particular to each simulation. All code is available on Github \citep{https://doi.org/10.5281/zenodo.18414840}. 

\begin{enumerate}
  \item\label{simu-case1} To study divergence from proportional aggregation, we fix the mean of the Gaussian ACID by $(d_1,d_2) = (2,-1)$. Per sample size $n$, we simulate data from SCM~\eqref{scm:generalcase} for varying values of $\beta_1 \in [-1,4]$. The results of the simulation are given in the plot in Figure~\ref{fig:sim_mismatch}A, which follows $\beta_1$ on the $x$-axis. The $y$-axis follows estimates of the causal effect of $A$ on $Y$: the IV estimates (computed by 2SLS) are plotted alongside the aggregate causal effect $\ACE(A,Y) = 2\beta_1-2$ and the theoretical IV estimand $\betaIV = \frac{\beta_1}{2}+1$.

  Proportional aggregation is achieved for $\beta_1 = 2$, for which $\ACE(A,Y) = \betaIV = 2$. We can see from the plot that as the value of $|\beta_1 - 2|$ grows, so does the difference between the $\ACE(A,Y)$ and the IV estimates. This difference can be arbitrarily large depending on the value of $\beta_1$. 
  \item\label{simi-case2}  To study the mismatch between the aggregate causal effect under varied Gaussian ACIDs and the IV estimates from observational data given by SCM~\eqref{scm:generalcase}, we fix $\beta_1 = 1$ and vary $d_1 \in [-2,2]$, which in turn also determines the values of $d_2$ according to the constraint in~\eqref{condnorm:mean}. For each sample size $n$, we again simulate data from SCM~\eqref{scm:generalcase}.
  The results of the simulation are given in the plot in Figure~\ref{fig:sim_mismatch}B, which follows $d_1$ on the $x$-axis. The $y$-axis follows estimates of the causal effect of $A$ on $Y$, including the sample IV estimates, the aggregate causal effect $\ACE(A,Y) =  2-d_1$, and the theoretical IV estimand $\betaIV = 1.5$. 
  
  The instrument-tuned intervention is achieved for $d_1 = 0.5$, for which $\ACE(A,Y) = \betaIV = 1.5$. As we move further away from the instrument-tuned intervention, that is, as $|d_1 - 0.5|$ grows, so does the difference between the $\ACE(A,Y)$ and the IV estimates. 
\end{enumerate}

\section{The Exclusion Assumption and its Justifications}\label{sec:diagnosis}

We have shown that unless proportional aggregation is satisfied or a valid ACID suitable for an instrument-tuned intervention is provided, the difference between the IV estimate of the causal effect and the aggregate causal effect is unbounded. What are the implications for IV studies that involve aggregate treatment variables?

It is tempting to see the aggregation problem as another variation of the ways in which the exclusion restriction
can be violated, and to propose the corresponding remedies. As we show in this section, the standard remedies do not carry over. The exclusion restriction requires that the instrument $I$ is independent of the outcome $Y$ given the treatment $A$ and confounder $U$. In our setting, while $I \ind Y| A_1, ..., A_k, U$, the exclusion restriction is indeed violated with respect to the aggregate treatment, since we have that $I \not\ind Y| A, U$ for $A = \sum_j \alpha_jA_j$ with arbitrary $\alpha_j$ unless proportional aggregation holds. This is a familiar problem that can also occur when a continuous treatment variable is discretized. Moreover, the Gaussian observational joint distribution over $I, U, A, Y$ in the aggregation setting of Figure~\ref{fig:case11} (SCM~\eqref{scm:generalcase}) is equivalent to the Gaussian observational joint distribution of the causal system in Figure~\ref{fig:er} (SCM~\eqref{scm:classic_erv_IV} below) that violates the exclusion restriction by having an extra edge for a direct effect of the instrument on the outcome, as reflected by the following result. See Supplementary Section~\ref{apx:aggexclproof} for the proof. 

\begin{figure}
  \centering  
\includegraphics[width = 7cm]{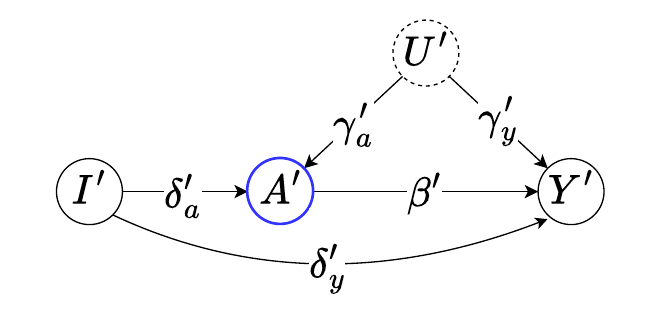}
\caption{A violation of the exclusion restriction that is equivalent to the aggregate setting under linearity and Gaussianity.}
 \label{fig:er}
\end{figure}

\begin{proposition}[Aggregation as an exclusion violation]\label{prop:agg_er_equiv}
Consider SCM~\eqref{scm:generalcase}, where the errors are assumed to be mutually independent standard Gaussians. The distribution of the vector $(I,U,A,Y)^{\top}$ is then equivalent to the distribution of the vector $(I',U',A',Y')^{\top}$ produced by the following 
linear SCM (compatible with the graph in Figure \ref{fig:er}): 
   \begin{align}
    U' &\leftarrow \epsilon_u'; \quad
    I' \leftarrow  \epsilon_i'  \\ 
    A' &\leftarrow \delta_a' I' + \gamma_{a}'U' + \epsilon_{a}' \label{scm:classic_erv_IV} \\ 
    Y' &\leftarrow \beta' A' + \gamma_{y}'U' + \delta_y' I' + \epsilon_y' 
\end{align}
where error terms are independent mean zero Gaussian with the following variances: 
\begin{align}
    \var(\epsilon_u') =\var(\epsilon_i')=1, \quad \var(\epsilon_a') = \sum_j \alpha_j^2, \quad \var(\epsilon_y') = 1 + \frac{\sum_{l<j}(\beta_l\alpha_j - \beta_j\alpha_l)^2}{\sum_j \alpha_j^2}.
\end{align} 
\end{proposition}

In the case of proportional aggregation (where $\beta_j/\alpha_j = \tau$, for $j \in \{1, \dots k\}$), the exclusion restriction is satisfied because in Proposition \ref{prop:agg_er_equiv} the coefficient on the direct effect from the instrument to the outcome is zero ($\delta_y' = 0$, see Supplementary Section~\ref{apx:aggexclproof}). 

{More generally, $\delta_y' = 0$ holds precisely when the aggregation coefficients satisfy
$$\big(\sum_j \alpha_j^2\big)\big(\sum_j \beta_j\delta_j\big) = \big(\sum_j \alpha_j\beta_j\big)\big(\sum_j \alpha_j\delta_j\big),$$
a condition of which proportional aggregation is one special case (see Supplementary Section~\ref{apx:aggexclproof}). This condition defines a measure-zero set of coefficients, so for almost all choices of $\alpha_j, \beta_j, \delta_j$ we have $\delta_y' \neq 0$ and the exclusion restriction is violated. We stress that the claim is not that aggregation \emph{always} induces a violation, rather, that an analyst with access only to the aggregate treatment $A$ cannot rule out $\delta_y' \neq 0$ without additional, and typically untestable, assumptions on the component coefficients.} The ACIDs derived in Section~\ref{sec:instaggeq} are just chosen in such a way that the IV estimand coincides with the aggregate causal effect. 

This point highlights the crux of considering existing remedies that address the exclusion restriction violations in other contexts: \emph{The adjustment that is needed to ensure that the IV estimand matches the aggregate causal effect depends on the particular ACID used to define the causal effect}. 
Thus, although extensive work addresses  exclusion restriction violations in linear SCMs, such as those from treatment measurement error, instrument endogeneity, or direct instrument effects \citep{vansteelandt2009correcting,zhu2022causal,biener2024non,cinelli2025omitted,ye2023instrumented}, these methods do not carry over here unless they coincide with a suitable ACID. 
Put differently, the $\beta'$ of Proposition \ref{prop:agg_er_equiv} will not, in general, equal the aggregate causal effect unless a particular ACID is assumed and justified.

Researchers often make a significant effort to provide domain-specific justifications for why they believe the exclusion restriction 
is satisfied in their application. However, a discussion of aggregate treatments does not generally appear in these justificatory arguments. 
In the over-identified case (when there are more instruments than treatments), practitioners often apply what is known as the ``Sargan test'' to argue that the exclusion restriction is satisfied \citep{sargan1958estimation}. The null hypothesis of the Sargan test is that the IV estimands corresponding to different instruments are equal  \citep{windmeijer2019two}. 
While rejecting the null hypothesis indicates a violation of the exclusion restriction, {failing to reject the null hypothesis does not directly imply that the exclusion restriction is satisfied \citep{kiviet2017discriminating}.} In the next section, we explore how one might use the Sargan test to check for proportional aggregation, and what additional justifications are needed.

\subsection{Using the Sargan Test to Test for Proportional Aggregation}

If there are at least two instruments, each standing in a configuration with $A_1, \dots, A_k$ and $Y$ as in the bottom of Figure~\ref{fig:case11}, then, under proportional aggregation, 
the IV estimand and resulting estimates should not be sensitive to the choice of instrument. 
Hence, under linearity and homoskedasticity, the null hypothesis of the Sargan test is satisfied under proportional aggregation  \citep{sargan1958estimation, windmeijer2019two}. 
Here we investigate the extent of this test's use for checking proportional aggregation in regards to instrument strength and the stand-alone implications of the null hypothesis.

\begin{table}
\centering
\parbox{110mm}{\caption{Instrument configurations and relevance with $A$} \label{table:instrument_strength}} \\
\fbox{%
\begin{tabular}{*{5}{c}}
 \emph{Instrument configuration} & $(\delta_{11}, \delta_{12})$  & $(\delta_{21}, \delta_{22})$ & $\text{cor}(I_1, A)$ & $\text{cor}(I_2, A)$\\ 
 \hline \rule{0pt}{3.5ex}
 Strong-Weak & $(5, 3)$ & $(0.1, 0.2)$ & $0.977$ & $0.037$ \\[1ex]
 Strong-Strong & $(5, 3)$ & $(4, 2)$ & $0.788$ & $0.591$ \\[1ex]
 Weak-Weak & $(0.15, 0.1)$ & $(0.08, 0.05)$ & $0.143$ & $0.074$ \\[1ex]
\end{tabular}}
\end{table}

\begin{figure}
\centering
\includegraphics[width = \linewidth]{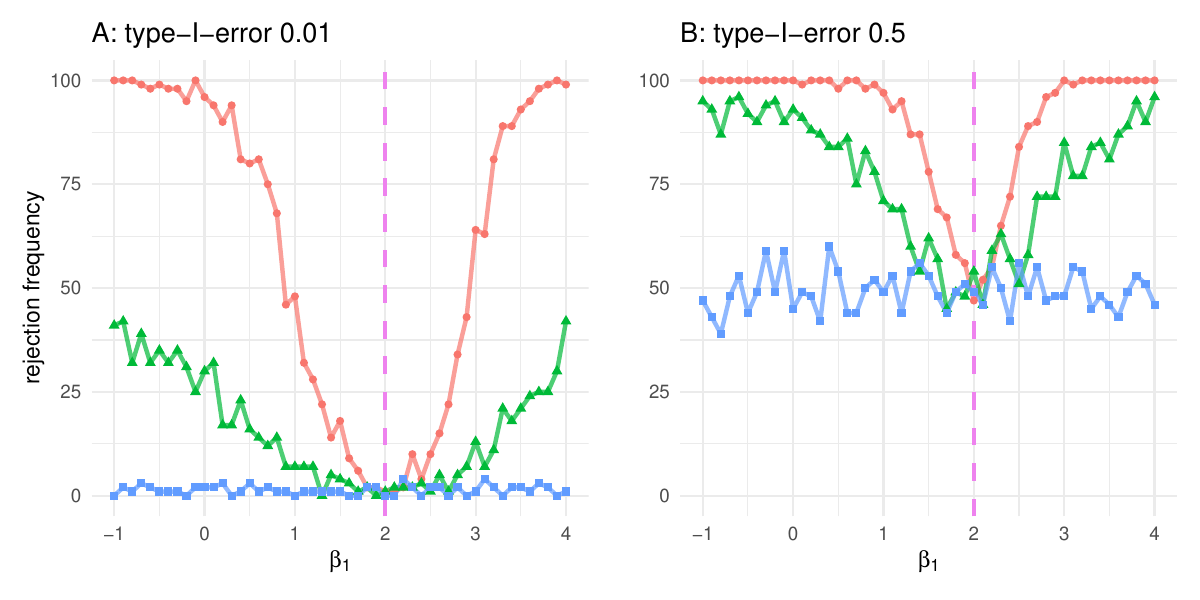}
\caption{Empirical power of the Sargan test for two strong instruments (red circles), one strong and one weak instrument (green triangles), and two weak instruments (blue squares) as a function of deviations from proportional aggregation for A: type-I-error $0.01$, B: type-I-error $0.5$. $H_0$ is true for  $\beta_1=2$ (marked with the purple dashed line).}
\label{fig:signf01_5}
\end{figure}

In a simulation we  illustrate the power of the Sargan test to check the assumption of proportional aggregation using two instruments $I_1$ and $I_2$ with varying strengths. 
We consider data generated by the following  SCM:
\begin{subequations}
\label{eq:sargan0}
\begin{align}
 &U \leftarrow \epsilon_u; \quad I_1 \leftarrow \epsilon_{i_1}; \quad I_2 \leftarrow \epsilon_{i_2} \\
  &A_1 \leftarrow  \delta_{11}I_1 + \delta_{21}I_2 + 0.5U + \epsilon_{a_1} \\
  &A_2 \leftarrow  \delta_{12}I_1 + \delta_{22}I_2 + 0.5U + \epsilon_{a_2}\\
  &A = A_1 + A_2\\
  &Y \leftarrow \beta_1 A_1 + 2 A_2 + 2U + \epsilon_y.
\end{align}
\end{subequations}
All error terms are mutually independent Gaussians with mean zero and variance 1. 
\cite{kiviet2021instrument} report that the power of the Sargan test is compromised when at least one of the two instruments is weak. We therefore consider three instrument configurations as described in Table~\ref{table:instrument_strength}: one where both instruments are weak, one where only one is weak, and one where both are strong.
Several formal definitions for a ``weak" instrument have been proposed (i.e.~the threshold for violation of the relevance assumption), with the rule-of-thumb for a weak instrument based on data being a first-stage F-statistic $<10$ \citep{stock2002survey}. Here, we use instrument-treatment  population correlations to {classify} an instrument as strong or weak. In particular, we say an instrument is strong if this correlation is above 0.5, and weak if the correlation is below 0.2. Our definition of strong and weak instruments is conservative with respect to the rule-of-thumb F-statistic $< 10$, as most of the datasets in even our two weak instruments setting have an observed F-statistic $\ge 10$ (see  Supplementary Section~\ref{app:sargan-fig} for details).

For each of the three instrument configurations, we simulate $100$ datasets each of sample size $n = 1000$ for each $\beta_1 \in [-1, 4]$ in one-tenth increments. Note that proportional aggregation is satisfied at $\beta_1 = 2$. For $\beta_1 \neq 2$, the null hypothesis of the Sargan test is false, and we would expect to reject the null hypothesis. For each of our settings we conduct a Sargan test with type-I-error $0.01$ and $0.5$ (similar to \citealp{kiviet2021instrument}). The rejection frequency for the Sargan tests with type-I-error $0.01$ is reported in Figure~\ref{fig:signf01_5}A, while Figure~\ref{fig:signf01_5}B  shows the rejection frequency with type-I-error $0.5$.  As $\beta_1$ shifts away from proportional aggregation, the null hypothesis of the Sargan test is more severely violated, so we expect to see a larger rejection frequency (indicating larger power of the test) in both plots.  

In the configuration with two strong instruments the test behaves as expected. However, when even one of the instruments is weak the Sargan test exhibits very low power when the type-I-error is $0.01$. This is somewhat corrected for the configuration with one strong instrument for the type-I-error $0.5$, but the configuration with two weak instruments has practically zero power for either choice of type-I-error. A similar pattern in the non-aggregate treatment setting was noted by  \cite{kiviet2021instrument}.
Based on the results of this simulation, we would recommend using a large type-I-error in combination with strong instruments for the Sargan test as a diagnostic tool.

Importantly, however, even if the null hypothesis of the Sargan test is true, meaning that the IV estimands obtained by different instruments are equal, that does not necessarily imply that proportional aggregation is satisfied. 
For instance, suppose we have $m$ instruments $I_1, \ldots, I_m$ such that the components $A_1,\dots, A_k$ are functions of all $m$ instruments. Let $\delta_{lj}$ denote the effect of $I_l$ on component variable $A_j$, $l \in \{1, \dots, m \}, j \in \{1, \dots, k\}$. The instruments are not correlated with $U$ and do not have an effect on $Y$ except through $A_1, \dots, A_k$.

Here the null hypothesis of the Sargan test is equivalent to the IV estimand $\betaIV^{I_b}$ obtained using instrument $I_b$ being equal to the IV estimand $\betaIV^{I_c}$ obtained using instrument $I_c$ for all pairs of instruments $I_b$ and $I_c$,  $b, c \in \{1,...,m\}$. Then,  \eqref{eq:2sls} implies: 
\begin{align}\label{eq:sargan1}
    \betaIV^{I_b} =\frac{\sum_j \beta_j \delta_{bj}}{{\sum_j \alpha_j \delta_{bj}}}
    = \frac{\sum_j \beta_j \delta_{cj}}{{\sum_j \alpha_j \delta_{cj}}} =     \betaIV^{I_c}.
\end{align}
Rearranging, we have that  \eqref{eq:sargan1} is satisfied as long as:
\begin{align}\label{eq:sargan2}
\left(\sum_j \beta_j \delta_{bj}\right)\left(\sum_j \alpha_j \delta_{cj}\right)
    = \left(\sum_j \beta_j \delta_{cj}\right)\left({{\sum_j \alpha_j \delta_{bj}}}\right).
\end{align}
Equation \eqref{eq:sargan2} is satisfied if, for example,
\begin{align}\label{eq:sargan3}
\sum_{i,j}(\beta_i\alpha_j - \alpha_i\beta_j)\delta_{bi}\delta_{cj} = 0
\text{\enspace\quad or \enspace\quad}
\sum_{i,j}\beta_i\alpha_j(\delta_{bi}\delta_{cj} - \delta_{ci}\delta_{bj}) = 0.
\end{align}
In the setting of proportional aggregation, we have that $\beta_i\alpha_j = \beta_j \alpha_i$ for all $i,j$, meaning that the first equality in \eqref{eq:sargan3} holds. However, there are many other settings that could lead to \eqref{eq:sargan2} being satisfied;
for instance, as implied by the second equality in \eqref{eq:sargan3}, if $\delta_{bj} = \delta_{cj}$ for all $j \in \{1, \dots, k\}$. Since any of these situations are possible under the null hypothesis of  the Sargan test, the justification of proportional aggregation requires more than just reporting a belief about the null hypothesis being true.

\section{Variations and Implications for Practice}\label{sec:extensions}

The aggregate setting displayed in the bottom of Figure~\ref{fig:case11}, corresponding to SCM \eqref{scm:generalcase}, 
is a generic case, but there are several sub-cases that are likely to be of interest in practice. For simplicity of illustration we only consider the aggregate $A$ to consist of two components $A_1, A_2$ in this section. Consider the following cases: 
(b) the entire treatment is instrumented, but only part of the treatment has an effect on the outcome, that is, $\beta_2 = 0$, (c) only part of the treatment is instrumented ($\delta_2 = 0$) but the entire treatment has an effect on the outcome, and (d) the treatment contains components that are irrelevant to instrument and outcome ($\delta_2 = \beta_2 = 0$). The IV estimand for these settings is given in the first row of Table~\ref{table:2sls}. As discussed before, unless one assumes proportional aggregation, the true aggregate causal effect depends on the specific ACID used. Under proportional aggregation, the aggregate causal effect is always given by $\tau$, which the estimands in cases (b)-(d) simplify to as well. That is, under proportional aggregation, the IV estimand matches the causal effect for these settings. 

For the 
Gaussian ACID discussed in Section~\ref{subsec:gaussian_example}, 
the aggregate causal effect is given in the second row of Table~\ref{table:2sls}. Unsurprisingly, the requirements for the IV estimand to match the aggregate causal effect are versions of the constraints in Section~\ref{sec:instaggeq} adjusted to the respective settings here. Overall, even though cases (b)-(d) represent scenarios that are likely to occur in the social sciences where IV techniques are used, the point is again that unless proportional aggregation holds, highly contrived ACIDs are needed for the IV estimand to match the aggregate causal effect.

\begin{table}
\centering
\parbox{100mm}{\caption{IV estimands and aggregate causal effects for settings 
described in Section \ref{sec:extensions} under a Gaussian ACID} \label{table:2sls}} \\
\fbox{%
\begin{tabular}{*{5}{c}}
 \emph{Case} & general & (b) $\beta_2 = 0$ & (c) $\delta_2 = 0$  & (d) $\delta_2 = \beta_2 = 0$\\ 
 \hline \rule{0pt}{5ex}
 {$\betaIV$} & \Large$\frac{ \beta_1\delta_1 + \beta_2\delta_2}{\alpha_1\delta_1+\alpha_2 \delta_2}$ 
 & \Large$\frac{\beta_1\delta_1}{\alpha_1\delta_1+\alpha_2 \delta_2}$ 
 & \Large$\frac{\beta_1}{\alpha_1}$ 
 & \Large$\frac{\beta_1}{\alpha_1}$ \\[1ex]
{$\ACE(A,Y)$} & $\beta_1d_1+ \beta_2d_2$
 &  $\beta_1d_1$
 & $\beta_1d_1+ \beta_2d_2$
 &  $\beta_1d_1$\\[1ex]
\end{tabular}}
\end{table}

In Supplementary Section~\ref{appendix:extensions} we also consider the possibility of aggregated instruments or aggregated outcomes. These are widely used \citep{spiga2023tools,lal2024much}, but as long as the treatment is not aggregated, then aggregated instruments or outcomes pose no problems for IV estimation. Similarly, the aggregation of confounders is not an issue since the adjustment set is irrelevant in an IV analysis.

The classic IV setting considers linear causal relations with additive Gaussian noise.
It is natural to ask what would happen in the absence of such restrictions.
The first thing to note is that without linearity and Gaussianity we generally lose  value independence (see Supplementary Section \ref{apx:counter_example}), that is, the true target causal effect is no longer well-defined for continuous treatment $A$, but will depend on the particular values $\opdo(A=a)$ that the aggregate $A$ is set to. Moreover, non-linearities in the causal effects will in general mean that the aggregate causal effect is going to become far more sensitive to how the aggregate value is realized through the ACID in the components. So, the consideration of non-linear causal effects does not appear to be a promising route to achieve a match between the IV estimand and the aggregate causal effect. However, as in other causal discovery settings, non-linearities may help identify the components of the treatment, and, once the components are identified, the causal effects can be estimated for each component individually. 

This last point leads to a more general point about what our results imply for handling the aggregate setting in practice. If proportional aggregation cannot reasonably be assumed, then an understanding of the aggregate-component relation becomes crucial: Is it possible to identify the components and estimate their effects separately? Is it possible to instantiate a specific ACID? Or is it possible to change the aggregation rule such that proportional aggregation can be achieved? Each of these possibilities takes us beyond the setting we consider here, where the components $A_1, \ldots, A_k$ are not known to the researcher. 

What might one do when facing aggregation in IV studies? The Sargan test offers a first, though insufficient, check for proportional aggregation. Our simulations in Section 4.3 show that if the component effects can be bounded near proportional aggregation, the aggregate causal effect can be bounded as well. Alternatives such as reinterpreting the estimand as a summary of per-component effects \citep{harris2022interpreting} or targeting a different estimand \citep{caetano2025causal} either yield a mixture of causal effects that is hard to interpret for intervention {or represent a special case of the $\ACE(A,Y)$}, and neither addresses the linear IV setting directly, though \cite{caetano2025causal} make conjectures for the IV case. Our message is different: rather than proposing a new estimand, we argue that for IV settings, researchers should carefully assess whether the treatment is aggregate in nature and, if so, the aggregation assumptions themselves must be made explicit and defended. Concretely, an aggregate treatment IV estimate retains its usual causal interpretation only if  proportional aggregation or a specific (instrument-tuned) intervention holds, and each of these should be justified alongside the exclusion restriction. The extension of our results to the discrete case, to which IVs are also widely applied, is of interest for future work.

\section{Conclusion}\label{sec:conclusion}

Instrumental variable analyses have long served as a compass for social scientists, economists, and epidemiologists, providing orientation in settings with unobserved confounding where causal identification would otherwise be elusive. Despite sustained scrutiny of instrument validity, the conditions of relevance, exclusion, and exchangeability offer a clear causal framework that has enabled impactful policy interventions, ranging from education and labour-market reforms to public health interventions, that would otherwise be infeasible using standard regression analysis alone.

However, when treatments are aggregates of multiple components that have different effects on the outcome, our results show that classical IV methods identify causal effects only under highly restrictive assumptions about the system or about how the instrument relates to the aggregation. We have described these assumptions as \emph{proportional aggregation} (Section~\ref{sec:propagg}) and \emph{instrument-tuned interventions} (Section~\ref{sec:instaggeq}).  On the one hand, these findings raise concerns about whether IV analyses can continue to provide clear directional guidance in settings with aggregated treatments, a common feature of social science applications. In such cases, aggregation in the linear Gaussian case cannot be distinguished from a violation of the exclusion restriction (Section~\ref{sec:diagnosis}), rendering statistical testing infeasible without additional assumptions. On the other hand, our results also provide guidance for analyses that may have become ``lost in aggregation'' by emphasizing that any claim that the exclusion restriction holds must not only rule out direct effects of the instrument on the outcome, but also justify why the treatment should not be viewed as an aggregate; alternatively, assumptions such as proportional aggregation or instrument-tuned interventions must be made explicit and defended.

These assumptions are strong and reflect fundamental limitations of the IV setting; in the absence of additional evidence, IV estimates should therefore be used with caution as guides for policy intervention when the treatment is plausibly an aggregate.

 \section*{Data availability}

  All results in this paper are based on simulated data. The code that
  generates the data and reproduces every numerical result, figure and table is publicly available at \url{https://github.com/danielletsao/lost-in-aggregation} and archived at  \url{https://doi.org/10.5281/zenodo.18414840}. No empirical datasets were
  analysed.

\section*{Acknowledgments}
We are very grateful for the detailed and constructive comments by two anonymous reviewers that resulted in a much improved manuscript. 
This work originated at the 2023 Causality Workshop at the Bellairs Research Institute that FE, KM \& EP attended.
FE would like to thank the participants of the 2025 Causality Workshop at Duke for their feedback on an initial presentation of the material, in particular Matt Masten, Betsy Ogburn and Jiji Zhang. DT, FE \& EP would like to thank the Isaac Newton Institute for Mathematical Sciences, Cambridge, for support and hospitality during the programme on Causality, where work on this paper was completed.

\section*{Funding}
This material is based upon work supported by the National Science Foundation under Grant No.\ 2210210, and the Natural Sciences and Engineering Research Council of Canada under Grant No.\ 599253.

\section*{Disclosure Statement}
The authors declare that they have no conflict of interest.

\bibliographystyle{plainnat}
\bibliography{ref.bib}

@article{harris2022interpreting,
  title={Interpreting instrumental variable estimands with unobserved treatment heterogeneity: The effects of college education},
  author={Harris, Clint},
  journal={arXiv preprint arXiv:2211.13132},
  year={2022}
}

@unpublished{caetano2025causal,
  title={Causal Inference for Aggregated Treatment},
  author={Caetano, Carolina and Caetano, Gregorio and Callaway, Brantly and Dyal, Derek},
  note={arXiv preprint arXiv:2506.22885},
  year={2025}
}

@article{vanderweele2009concerning,
  title={Concerning the consistency assumption in causal inference},
  author={VanderWeele, Tyler J},
  journal={Epidemiology},
  volume={20},
  number={6},
  pages={880--883},
  year={2009},
  publisher={LWW}
}

@article{imbens1994identification,
  title={Identification and Estimation of Local Average Treatment Effects},
  author={Imbens, Guido W and Angrist, Joshua D},
  journal={Econometrica},
  volume={62},
  number={2},
  pages={467--475},
  year={1994}
}

@article{iong2024latent,
  title={A latent mixture model for heterogeneous causal mechanisms in {M}endelian randomization},
  author={Iong, Daniel and Zhao, Qingyuan and Chen, Yang},
  journal={The Annals of Applied Statistics},
  volume={18},
  number={2},
  pages={966--990},
  year={2024}
}

@article{theil1953repeated,
  title={Repeated least squares applied to complete equation systems},
  author={Theil, Henri},
  journal={The Hague: Central Planning Bureau},
  year={1953}
}

@article{basmann1957generalized,
  title={A generalized classical method of linear estimation of coefficients in a structural equation},
  author={Basmann, Robert L},
  journal={Econometrica: Journal of the Econometric Society},
  pages={77--83},
  year={1957}
}

@book{hernancausal,
  title={Causal Inference: What If},
  author={Hernan, Miguel A and Robins, James M},
year={2026},
  publisher={Taylor \& Francis}
}

@book{pearl2009causality,
  title={Causality},
  author={Pearl, Judea},
  year={2009},
  publisher={Cambridge {U}niversity {P}ress}
}

@article{goldsmith2020bartik,
  title={Bartik instruments: What, when, why, and how},
  author={Goldsmith-Pinkham, Paul and Sorkin, Isaac and Swift, Henry},
  journal={American Economic Review},
  volume={110},
  number={8},
  pages={2586--2624},
  year={2020},
  publisher={American Economic Association 2014 Broadway, Suite 305, Nashville, TN 37203}
}

@article{vansteelandt2009correcting,
  title={Correcting instrumental variables estimators for systematic measurement error},
  author={Vansteelandt, Stijn and Babanezhad, Manoochehr and Goetghebeur, Els},
  journal={Statistica Sinica},
  volume={19},
  pages={1223},
  year={2009}
}

@article{angrist1991does,
  title={Does compulsory school attendance affect schooling and earnings?},
  author={Angrist, Joshua D and Krueger, Alan B},
  journal={The Quarterly Journal of Economics},
  volume={106},
  number={4},
  pages={979--1014},
  year={1991},
  publisher={MIT Press}
}

@article{bound1995problems,
  title={Problems with instrumental variables estimation when the correlation between the instruments and the endogenous explanatory variable is weak},
  author={Bound, John and Jaeger, David A and Baker, Regina M},
  journal={Journal of the American Statistical Association},
  volume={90},
  number={430},
  pages={443--450},
  year={1995},
  publisher={Taylor \& Francis}
}

@article{buckles2013season,
  title={Season of birth and later outcomes: Old questions, new answers},
  author={Buckles, Kasey S and Hungerman, Daniel M},
  journal={Review of Economics and Statistics},
  volume={95},
  number={3},
  pages={711--724},
  year={2013},
  publisher={The MIT Press}
}

@article{vanderweele2013causal,
  title={Causal inference under multiple versions of treatment},
  author={VanderWeele, Tyler J and Hernan, Miguel A},
  journal={Journal of Causal Inference},
  volume={1},
  number={1},
  pages={1--20},
  year={2013},
  publisher={De Gruyter}
}

@article{nordestgaard2012effect,
  title={The effect of elevated body mass index on ischemic heart disease risk: {C}ausal estimates from a {M}endelian randomisation approach},
  author={Nordestgaard, B{\o}rge G and Palmer, Tom M and Benn, Marianne and Zacho, Jeppe and Tybjaerg-Hansen, Anne and Davey Smith, George and Timpson, Nicholas J},
  journal={PLOS Medicine},
  volume={9},
  number={5},
  year={2012},
  publisher={Public Library of Science San Francisco, USA}
}

@article{rubin1986comment,
  title={Comment: Which ifs have causal answers},
  author={Rubin, Donald B},
  journal={Journal of the American Statistical Association},
  volume={81},
  number={396},
  pages={961--962},
  year={1986},
  publisher={Taylor \& Francis}
}

@inproceedings{beckers2020approximate,
  title={Approximate causal abstractions},
  author={Beckers, Sander and Eberhardt, Frederick and Halpern, Joseph Y},
  booktitle={Proceedings of the Conference on Uncertainty in Artificial Intelligence},
  pages={606--615},
  year={2020}
}

@article{spirtes2004causal,
  title={Causal inference of ambiguous manipulations},
  author={Spirtes, Peter and Scheines, Richard},
  journal={Philosophy of Science},
  volume={71},
  number={5},
  pages={833--845},
  year={2004},
  publisher={Cambridge University Press}
}

@article{gabriel2024impact,
  title={The impact of coarsening an exposure on partial identifiability in instrumental variable settings},
  author={Gabriel, Erin E and Sachs, Michael C and Sj{\"o}lander, Arvid},
  journal={Biostatistics},
  volume={26},
  year={2025},
  publisher={Oxford University Press}
}

@inproceedings{zhu2024meaningful,
  title={Meaningful causal aggregation and paradoxical confounding},
  author={Zhu, Yuchen and Budhathoki, Kailash and K{\"u}bler, Jonas M and Janzing, Dominik},
  booktitle={Proceedings of Causal Learning and Reasoning},
  pages={1192--1217},
  year={2024}
}

@inproceedings{zhu2022causal,
  title={Causal inference with treatment measurement error: {A} nonparametric instrumental variable approach},
  author={Zhu, Yuchen and Gultchin, Limor and Gretton, Arthur and Kusner, Matt J and Silva, Ricardo},
  booktitle={Proceedings of the Conference on Uncertainty in Artificial Intelligence},
  pages={2414--2424},
  year={2022}
}

@article{durbin1954errors,
  title={Errors in variables},
  author={Durbin, James},
  journal={Revue de l'{I}nstitut International de Statistique},
  pages={23--32},
  year={1954},
  publisher={JSTOR}
}

@article{burauel2023evaluating,
  title={Evaluating instrument validity using the principle of independent mechanisms},
  author={Burauel, Patrick F},
  journal={Journal of Machine Learning Research},
  volume={24},
  number={176},
  pages={1--56},
  year={2023}
}

@article{kitagawa2015test,
  title={A test for instrument validity},
  author={Kitagawa, Toru},
  journal={Econometrica},
  volume={83},
  number={5},
  pages={2043--2063},
  year={2015},
  publisher={Wiley Online Library}
}

@article{burgess2017review,
  title={A review of instrumental variable estimators for {M}endelian randomization},
  author={Burgess, Stephen and Small, Dylan S and Thompson, Simon G},
  journal={Statistical Methods in Medical Research},
  volume={26},
  number={5},
  pages={2333--2355},
  year={2017},
  publisher={SAGE Publications Sage UK: London, England}
}

@article{angrist1996identification,
  title={Identification of causal effects using instrumental variables},
  author={Angrist, Joshua D and Imbens, Guido W and Rubin, Donald B},
  journal={Journal of the American Statistical Association},
  volume={91},
  number={434},
  pages={444--455},
  year={1996},
  publisher={Taylor \& Francis}
}

@article{acemoglu2008income,
  title={Income and democracy},
  author={Acemoglu, Daron and Johnson, Simon and Robinson, James A and Yared, Pierre},
  journal={American Economic Review},
  volume={98},
  number={3},
  pages={808--842},
  year={2008},
  publisher={American Economic Association}
}

@article{rashad2006structural,
  title={Structural estimation of caloric intake, exercise, smoking, and obesity},
  author={Rashad, Inas},
  journal={The Quarterly Review of Economics and Finance},
  volume={46},
  number={2},
  pages={268--283},
  year={2006},
  publisher={Elsevier}
}

@phdthesis{reiersol1945confluence,
  title={Confluence analysis by means of instrumental sets of variables},
  author={Reiers{\o}l, Olav},
  year={1945},
  school={Almqvist \& Wiksell}
}

@article{lal2024much,
  title={How Much Should We Trust Instrumental Variable Estimates in Political Science? {Practical} Advice Based on 67 Replicated Studies},
  author={Lal, Apoorva and others},
  journal={Political Analysis},
  pages={1--20},
  year={2024},
  publisher={Cambridge University Press}
}

@article{cinelli2025omitted,
  title={An omitted variable bias framework for sensitivity analysis of instrumental variables},
  author={Cinelli, Carlos and Hazlett, Chad},
  journal={Biometrika},
  volume={112},
  number={2},
  year={2025},
  publisher={Oxford University Press}
}

@article{ye2023instrumented,
  title={Instrumented difference-in-differences},
  author={Ye, Ting and Ertefaie, Ashkan and Flory, James and Hennessy, Sean and Small, Dylan S},
  journal={Biometrics},
  volume={79},
  number={2},
  pages={569--581},
  year={2023},
  publisher={Wiley Online Library}
}

@article{biener2024non,
  title={Non-classical measurement error in instrumental variables estimation: An application to the medical care costs of obesity},
  author={Biener, Adam I and Meyerhoefer, Chad and Cawley, John},
  journal={Health Economics},
  volume={33},
  number={11},
  pages={2558--2574},
  year={2024},
  publisher={Wiley Online Library}
}

@article{kiviet2021instrument,
  title={Instrument approval by the {S}argan test and its consequences for coefficient estimation},
  author={Kiviet, Jan F and Kripfganz, Sebastian},
  journal={Economics Letters},
  volume={205},
  pages={109935},
  year={2021},
  publisher={Elsevier}
}

@article{windmeijer2019two,
  title={Two-stage least squares as minimum distance},
  author={Windmeijer, Frank},
  journal={The Econometrics Journal},
  volume={22},
  number={1},
  pages={1--9},
  year={2019},
  publisher={Oxford University Press}
}

@article{sargan1958estimation,
  title={The estimation of economic relationships using instrumental variables},
  author={Sargan, John D},
  journal={Econometrica: Journal of the Econometric Society},
  pages={393--415},
  year={1958},
  publisher={JSTOR}
}

@article{stock2002survey,
  title={A survey of weak instruments and weak identification in generalized method of moments},
  author={Stock, James H and Wright, Jonathan H and Yogo, Motohiro},
  journal={Journal of Business \& Economic Statistics},
  volume={20},
  number={4},
  pages={518--529},
  year={2002},
  publisher={Taylor \& Francis}
}

@article{theil1959aggregation,
  title={The aggregation implications of identifiable structural macrorelations},
  author={Theil, Henri},
  journal={Econometrica: Journal of the Econometric Society},
  pages={14--29},
  year={1959},
  publisher={JSTOR}
}

@article{lichtenberg1990aggregation,
  title={Aggregation of variables in least-squares regression},
  author={Lichtenberg, Frank R},
  journal={The American Statistician},
  volume={44},
  number={2},
  pages={169--171},
  year={1990},
  publisher={Taylor \& Francis}
}

@article{spiga2023tools,
  title={Tools for assessing quality and risk of bias in {M}endelian randomization studies: a systematic review},
  author={Spiga, Francesca and Gibson, Mark and Dawson, Sarah and Tilling, Kate and Davey Smith, George and Munafo, Marcus R and Higgins, Julian PT},
  journal={International Journal of Epidemiology},
  volume={52},
  number={1},
  pages={227--249},
  year={2023}
}

@book{theil1954linear,
  title={Linear aggregation of economic relations},
  author={Theil, Henri},
  year={1954},
  publisher={North-Holland Publishing Company, Amsterdam}
}

@article{kiviet2017discriminating,
  title={Discriminating between (in)valid external instruments and (in)valid exclusion restrictions},
  author={Kiviet, Jan F},
  journal={Journal of Econometric Methods},
  volume={6},
  number={1},
  pages={20160005},
  year={2017},
  publisher={De Gruyter}
}

@misc{https://doi.org/10.5281/zenodo.18414840,
  doi = {10.5281/zenodo.18414841},
  url = {https://doi.org/10.5281/zenodo.18414840},
  author = {Danielle Tsao},
  title = {Lost {I}n {A}ggregation: {T}he {C}ausal {I}nterpretation of the {I}nstrumental {V}ariable {E}stimand, {G}ithub Repository},
  year = {2026}
}

\appendix

\section{Full IV Calculations}
 In classic (non-aggregate) 2SLS estimation, $A$ is first regressed on $I$ and the linear predictor $\tilde{A}$ from this regression is then used in the second stage. At the population level, 
$
\tilde{A} = \frac{\cov(A,I)}{\var(I)}I.
$

In the second stage, $Y$ is regressed on $\tilde{A}$. The IV estimand is the coefficient for $\tilde{A}$ from the second stage regression, which by relevance, the exclusion restriction, and exchangeability is simplified as follows:
\begin{align}
     \frac{\cov(Y, \tilde{A})}{\var(\tilde{A})}  
    = \frac{\cov\left(Y, \frac{\cov(A,I)}{\var(I)}I\right)}{\var\left(\frac{\cov(A,I)}{\var(I)}I\right)}  
    = \frac{\cov(Y,I)}{\cov(A,I)}, \label{eq:classic2sls}
\end{align}
such that the rightmost side is the causal effect of $A$ on $Y$. 

In contrast, for 2SLS in the aggregate setting (Figure~\ref{fig:case11}),  
the linear predictor $\tilde{A}$ from the first stage regression is now
\begin{align}
    \tilde{A} &= \frac{\cov(\sum_j \alpha_jA_j,I)}{\var(I)}I. 
\end{align}
In turn, the updated IV estimand, denoted here as $\betaIV$, is
\begin{align}
    \betaIV &= \frac{\cov(Y,\tilde{A})}{\var(\tilde{A})} = \frac{\sum_j \beta_j\delta_j}{\sum_j \alpha_j\delta_j} 
\end{align}
due to the following simplifications under SCM~\eqref{scm:generalcase}: 
\begin{align}
\cov(Y,\tilde A) \enspace &= \-\ \frac{\cov(\sum_j \alpha_jA_j,I)}{\var(I)} \cov(\sum_j \beta_jA_j + \gamma_{y}U + \epsilon_y, I) \\
&= \frac{\cov(\sum_j \alpha_jA_j,I)}{\var(I)} \sum \beta_j\cov(A_j, I) \label{eq:covYI}  \\
\var(\tilde A) &= \frac{\cov^2(\sum_j \alpha_jA_j,I)}{\var(I)} = \frac{\cov(\sum_j \alpha_jA_j,I)}{\var(I)} \sum_j \alpha_j\cov(A_j,I)\\ 
\cov(A_j, I) &= \-\ \cov(\delta_j I + \gamma_{a_j}U + \epsilon_{a_j}, I) = \delta_j\var(I) \label{eq:covA1I} 
\end{align}

\section{Examples of Valid and Invalid ACIDs}
\subsection{Surgicality and the Aggregation Rule Do Not Imply Value Independence}
\label{apx:counter_example}

We present an ACID that satisfies surgicality and the aggregation rule but not value independence under SCM~\eqref{scm:generalcase} with $k=2$ components. First, for independent variables $Z_1 \sim \text{Uniform}[-2,2]$ and $Z_2 \sim \text{Uniform}[-1,1]$, note that the unconditional joint density function $p_{Z_1, Z_2}$ is  
\begin{align}
    p_{Z_1, Z_2}(z_1, z_2) = \frac{1}{4}\times \frac{1}{2} = \frac{1}{8}, \text{ \enspace $(z_1,z_2) \in [-2,2] \times [-1,1]$}. 
\end{align}
The probability density function $p_Z$ of $Z = Z_1 + Z_2$ is then
    \begin{align}
    p_Z(a) &= \int_{-1}^{1} p_{Z_1}(a - z_2)p_{Z_2}(z_2)dz_2\\
    &= \int_{\max\{a-2, -1\}}^{\min\{a+2, 1\}} \frac{1}{8}dz_2 
    = 
    \begin{cases}
        \int_{-1}^{a+2} \frac{1}{8} dz_2 = \frac{a+3}{8}, &a \in [-3,-1] \\ 
        \int_{-1}^{1} \frac{1}{8} dz_2 = \frac{1}{4}, &a \in [-1,1] \\ 
        \int_{a-2}^{1} \frac{1}{8} dz_2 = \frac{3-a}{8}, &a \in [1,3],  
    \end{cases}
    \end{align}
    where $p_{Z_1}$ and $p_{Z_2}$ denote the marginal distributions of $Z_1$ and $Z_2$ respectively. 
    
  Then to construct the ACID $P^*$ of interest for $(A_1, A_2)^\top$,  
    consider $P^*$ where $A_1 = a - A_2$ and the marginal density of $A_2$ matches the conditional density function of $Z_2$ given that $Z = a$: 
    \begin{align}
    p_2^*(a_2 ; a) 
    &= \frac{1}{p_Z(a)} \times \frac{1}{8}, \enspace a_2 \in [\max\{a-2, -1\}, \min\{a+2,1\}] \\ 
    &= 
    \begin{cases}
        \frac{1}{a+3}, &a \in [-3,-1], \enspace a_2 \in [-1, a+2]\\ 
        \frac{1}{2}, &a \in [-1,1], \enspace a_2 \in [-1,1 ] \\ 
        \frac{1}{3-a}, &a \in [1,3], \enspace a_2 \in [a-2, 1].\\ 
    \end{cases}
     \end{align}
     
For simplicity, consider the SCM~\eqref{scm:generalcase} coefficients: $\beta_1 = 2$, $\beta_2 = 3$, and $\alpha_j = 1$, $\delta_j = \gamma_{a_j} = 0$ for $j =1,2$. 
Then we have under $P^*$ that $\E[A_1 \,|\, \opdo(A=a)] = a-\E[A_2 \,|\, \opdo(A=a)]$ and 
    \begin{align}
    \E[A_2 | \opdo(a)] &= 
    \int a_2 dP^*(a_2 ;a) 
    = 
        \begin{cases}
        \int_{-1}^{a+2} \frac{a_2}{a+3}da_2 =  \frac{a+1}{2}, &a \in [-3,-1]\\ 
        \int_{-1}^1 \frac{a_2}{2}da_2 = 0, &a \in [-1,1] \\ 
        \int_{a-2}^1 \frac{a_2}{3-a}da_2 = \frac{a-1}{2}, &a \in [1, 3]. 
        \end{cases}
    \end{align}
    To see that value independence does not hold, observe for 
    $a=-2$ that 
    \begin{align}
        \E[Y|\opdo(A = -1)] -  \E[Y|\opdo(A = -2)] = 
          \frac{5}{2}
    \end{align}
    whereas for $a=-0.9$, 
    \begin{align}
        \E[Y|\opdo(A = 0.1)] -  \E[Y|\opdo(A = -0.9)] =  2. 
    \end{align}

\subsection{Example of a Valid Non-Gaussian ACID}\label{apx:valid_ACIDs}

Consider an ACID that satisfies surgicality and the aggregation rule, such that the marginal distributions of $\alpha_1A_1, ..., \alpha_kA_k$ under the ACID are identical. Such an ACID would satisfy value independence. To see this, observe that $\sum_j \alpha_j\E[A_j | \opdo(A=a)] = a$  and $\alpha_i\E[A_i | \opdo(A=a)] = \alpha_j\E[A_j | \opdo(A=a)]$ for all $i, j$. Hence, $\E[A_j| \opdo(A = a)] = \frac{a}{k\alpha_j}$ for $j=1,...,k$. It follows that $\ACE(A,Y) = \sum_j\frac{\beta_j}{k\alpha_j}$, which does not depend on $a$.

\subsection{The ``Natural'' Gaussian ACID} \label{apx:naturalACD}

One may want to consider an ACID that (surgically) sets  the component variables $(A_1, ..., A_k)^\top$ to the values they would have in the observational distribution, conditional on the aggregate constraint $A=a$. For our linear Gaussian setting, this amounts to appropriately parametrizing the  
multivariate Gaussian distribution in \eqref{condnorm:distn} to match the conditional distribution of $(A_1, ..., A_k)^\top$ given the aggregation rule from their observational joint distribution in SCM~\eqref{scm:generalcase}. 

Let $\rho_{ij}$ denote the observational covariance of $A_i$ with $A_j$ for $i,j \in \{1,...,k\}$, let $\it{\Sigma}'$ denote the observational covariance matrix for $(A_1, 
\dots, A_k)^\top$, and let $\mathbf{\alpha} = (\alpha_1, \dots, \alpha_k)^{\top}$. Then under SCM~\eqref{scm:generalcase}, the ``Natural'' Gaussian ACID
\begin{align}
(A_1, \dots , A_k)^\top 
\sim 
\mathcal{N}\left(
(c_1+ a d_1 \dots \ c_k+ ad_k)^\top,
\mathit{\Sigma} 
\right) 
\end{align}
has the mean parameters $(c_1, 
\dots, c_k)^{\top} = \mathbf{0}$ and  
\begin{align}\label{naturalparam:mean}
(d_1, 
\dots, d_k)^{\top} = 
 \frac{\mathit{\Sigma'}
\mathbf{\alpha}}{\mathbf{\alpha}^{\top}\mathit{\Sigma'}{\mathbf{\alpha}}}
= \frac{1}{\sum_{i,j}\alpha_i\rho_{ij}\alpha_j } ( \sum_j \rho_{1j}\alpha_j, \dots, \sum_j \rho_{kj}\alpha_j)^\top,
\end{align}
and the covariance matrix
\begin{align}\label{naturalparam:cov}
\mathit{\Sigma} = \mathit{\Sigma'} - \frac{\mathit{\Sigma'}\mathbf{\alpha}\mathbf{\alpha^\top}\mathit{\Sigma'}}{\mathbf{\alpha}^\top \mathit{\Sigma'} \mathbf{\alpha}},
\end{align}
all of which are derived from conditioning the observational distribution from SCM~\eqref{scm:generalcase} on the aggregation rule. 
Here, the fractional constant is simply $1/\var(A)$ and the elements of the conditional mean vector are the covariances of the components with $A$. It follows that
\begin{align}\label{ate:naturalACD}
\ACE(A,Y) = \frac{\sum_j \beta_j\cov(A_j, A)}{\var(A)}. 
\end{align}
In general, this will not be equal to the IV estimand in \eqref{eq:2sls} unless proportional aggregation holds or an instrument-tuned intervention is assumed.

\section{Partially Instrument-tuned Intervention Calculations} \label{app:partial-iti}
Suppose that for $j = \{1, \dots, l \}$, $1 < l < k$, we have
 $\frac{\beta_j}{\alpha_j} = \tau$.

Rewriting $\sum_{j=1}^k \beta_jA_j$ in terms of $\tau$ and plugging into equation \eqref{eq:ATE} yields: 
\begin{align}\label{eq:ace-partial-agg}
    \ACE(A,Y) = \tau + \sum_{j=l+1}^k(\beta_j - \tau \alpha_j)d_j.
\end{align}
Furthermore, the IV estimand corresponds to: 
\begin{align}\label{eq:beta-iv-partial-agg}
    \betaIV  &= \frac{\sum_{j=1}^k \beta_j\delta_j}{\sum_{j=1}^k \alpha_j\delta_j} = \tau + \frac{\sum_{j=l+1}^k (\beta_j-\tau\alpha_j)\delta_j}{\sum_{j=1}^k \alpha_j\delta_j}.
\end{align}
Then, from \eqref{eq:ace-partial-agg} and \eqref{eq:beta-iv-partial-agg}, we have that $\betaIV = \ACE(A,Y)$ if the following equality holds:
\begin{align}\label{eq:partial-prop-agg}
\frac{\sum_{j=l+1}^k (\beta_j-\tau\alpha_j)\delta_j}{\sum_{j=1}^k \alpha_j\delta_j} =  \sum_{j=l+1}^k (\beta_j - \tau \alpha_j ) d_j,  
\end{align}
which is true if for instance, \eqref{eq:instr_surg_eq} holds for $d_j$, $j \in \{l+1, \dots, k\}$, while the remaining parameters $d_1, \dots, d_l$ and $c_1, \dots, c_k$ and $\Sigma$ need only be chosen such that the constraints \eqref{condnorm:mean} and \eqref{condnorm:sigma} are satisfied. Hence, for this setting, we are only instrument-tuning slopes of those components that do not obey proportional aggregation. Of course, such a partially instrument-tuned intervention needs to be justified by strong assumptions on the component relationships with the outcome and the aggregate.

\section{Propositions~\ref{prop:char} and \ref{prop:agg_er_equiv} proofs}
\label{apx:aggexclproof} 

\begin{proof}[Proposition~\ref{prop:char}]
By \eqref{eq:2sls}, $\betaIV = \sum_j \beta_j\delta_j \big/ \sum_l \alpha_l\delta_l = \sum_j \beta_j s_j$, where $\sum_j \alpha_j s_j = 1$. By \eqref{condnorm:mean}, the aggregation rule gives $\sum_j \alpha_j d_j = 1$. Hence,
\begin{align}
   \ACE(A,Y) - \betaIV
   &= \sum_j \beta_j(d_j - s_j)
    = \sum_j (\tau\alpha_j + r_j)(d_j - s_j) \\
   &= \tau\sum_j \alpha_j(d_j - s_j) + \sum_j r_j(d_j - s_j)
    = \sum_j r_j(d_j - s_j),
\end{align}
since $\sum_j \alpha_j d_j = \sum_j \alpha_j s_j = 1$. 
\end{proof}

\begin{proof}[Proposition~\ref{prop:agg_er_equiv}]
    In the Gaussian case, showing Markov equivalence of SCMs~\eqref{scm:generalcase} and \eqref{scm:classic_erv_IV} amounts to matching their implied covariance matrices.
    By definition, 
    \begin{align*}
        \var(I) = \var(U) = \var(I') = \var(U') = 1 \text{ and }  \cov(I,U) = \cov(I', U') = 0. 
    \end{align*}
    The remaining entries of the respective covariance matrices 
    are computed as follows. Beginning with the covariance matrix for $(I,U,A,Y)^{\top}$, it is helpful to expand the equations for $Y$ and $A$ from SCM~\eqref{scm:generalcase} with respect to the errors terms: 
    \begin{align}
        A &= \epsilon_i(\sum_j \alpha_j\delta_j) + \epsilon_u(\sum_j \alpha_j\gamma_{a_j}) + \sum_j \alpha_j\epsilon_{a_j} \\
        Y &\leftarrow \epsilon_i(\sum_j \beta_j\delta_j) + \epsilon_u(\sum_j \beta_j\gamma_{a_j} + \gamma_y) + \sum_j \beta_j\epsilon_{a_j} + \epsilon_y
    \end{align}
    Since all errors are mutually independent with unit variance, it follows that
    \begin{align}
        \cov(I, A) &=  
        \sum_j \alpha_j\delta_j,  \quad 
        \cov(U,A) = 
        \sum_j \alpha_j\gamma_{a_j}, \quad  
        \cov(I, Y) =  
        \sum_j \beta_j\delta_j \\
        \cov(U,Y) &= 
        \sum_j \beta_j\gamma_{a_j} + \gamma_y \label{sys_scm:classicIV} \\
        \cov(A,Y) &= 
        (\sum_j \alpha_j\delta_j)(\sum_j \beta_j\delta_j) + (\sum_j \alpha_j\gamma_{a_j})(\sum_j \beta_j\gamma_{a_j} + \gamma_y) + \sum_j \alpha_j\beta_j, \\
        \var(A) &= (\sum_j \alpha_j\delta_j)^2 + (\sum_j \alpha_j\gamma_{a_j})^2 + \sum_j \alpha_j^2\\ 
        \var(Y) &= (\sum_j \beta_j\delta_j)^2 + (\sum_j \beta_j\gamma_{a_j} + \gamma_y)^2 + \sum\beta_j^2 + 1 
    \end{align}
For the covariance matrix of $(I',U',A',Y')^{\top}$, $Y'$ and $A'$ can similarly be rewritten as
    \begin{align}
        A' &\leftarrow \delta'_a\epsilon_i' + \gamma'_a\epsilon_u' + \epsilon_a' \\
         Y' &\leftarrow \epsilon_i'(\beta'\delta'_a + \delta_y') + \epsilon_u'(\beta'\gamma'_a + \gamma_y') + \beta'\epsilon_a' + \epsilon_y'. 
    \end{align} 
    By the above, we then have, 
     \begin{align}
        \cov(I', A') &=  \delta_a', \quad   \cov(U',A') = 
        \gamma_a', \quad \cov(I', Y') =  
        \beta'\delta_a' + \delta_y'  \\
        \cov(U',Y') &= 
        \beta'\gamma_a' + \gamma_y' \label{sys_eq:ER_violation} \\
        \cov(A',Y') &= 
        \beta'{\delta_a'}^2 + \delta_y'\delta_a' + \beta'{\gamma_a'}^2 + \gamma_y'\gamma_a' + \beta'\var(\epsilon_a') \\
        \var(A') &= {\delta_a'}^2 + {\gamma_a'}^2 + \var(\epsilon_a'), \\ 
        \var(Y') &=  (\beta'\delta'_a + \delta_y')^2 + (\beta'\gamma'_a + \gamma_y')^2 + {\beta'}^2\var(\epsilon_a') + \var(\epsilon_y')
    \end{align}
    The covariances of the treatment with the instrument and the unobserved confounder are straightforward to match between \eqref{sys_scm:classicIV} and \eqref{sys_eq:ER_violation}, i.e. 
     $\cov(I, A) = \cov(I', A')$ and $\cov(U, A) = \cov(U', A')$ imply that
    \begin{align}\label{matching_simplif}
        \delta_a' = \sum_j \alpha_j\delta_j, \enspace  
        \gamma_a' = \sum_j \alpha_j\gamma_{a_j}.
    \end{align}
    Matching the covariances of the outcome with the instrument, confounder, and treatment and substituting in constraints from \eqref{matching_simplif} yields: 
    \begin{align}
        \sum_j \beta_j\delta_j = \beta'\delta_a' + \delta_y', \qquad 
        &\sum_j \beta_j\gamma_{a_j} + \gamma_y = \beta'\gamma_a' + \gamma_y' \label{matching:UY} \\ 
       \delta_a'(\sum_j \beta_j\delta_j) + \gamma_a'(\sum_j \beta_j\gamma_{a_j} + \gamma_y) &+ \sum_j \alpha_j\beta_j
        =  \\
        &\beta'{\delta_a'}^2 + \delta_y'\delta_a' + \beta'{\gamma_a'}^2 + \gamma_y'\gamma_a' + \beta'\var(\epsilon_a') \label{matching:AY}
    \end{align}
    Then \eqref{matching:UY} implies that
    \begin{align}
        \delta_y' &= \sum_j \beta_j\delta_j -  \beta'\delta_a', \quad
        \gamma_y' = \sum_j \beta_j\gamma_{a_j} + \gamma_y -  \beta'\gamma_a' \label{matching:gamma_y'}
    \end{align}
    and plugging into \eqref{matching:AY}, 
    \begin{align}
        \delta_a'(\beta'\delta_a' + \delta_y')& + \gamma_a'(\beta'\gamma_a' + \gamma_y') + \sum_j \alpha_j\beta_j\\
        &= \beta'{\delta_a'}^2 + \delta_y'\delta_a' + \beta'{\gamma_a'}^2 + \gamma_y'\gamma_a' + \beta'\var(\epsilon_a')
    \end{align}
    which simplifies to 
    \begin{align}
    \beta' = \frac{\sum_j \alpha_j\beta_j}{\var(\epsilon_a')}. 
    \end{align}
    Matching the variance of the treatment, we have
    \begin{align}
        (\sum_j \alpha_j\delta_j)^2 + (\sum_j \alpha_j\gamma_{a_j})^2 + \sum_j \alpha_j^2 &= {\delta_a'}^2 + {\gamma_a'}^2 + \var(\epsilon_a'), 
    \end{align}
    and plugging in the constraints from \eqref{matching_simplif} yields:
    \begin{align}
        \var(\epsilon_a') = \sum_j \alpha_j^2.  
    \end{align}
    Similarly, matching the variance of the outcome yields
    \begin{align}
        (\sum_j\beta_j\delta_j)^2 + &(\sum_j \beta_j\gamma_{a_j} + \gamma_y)^2 + \sum\beta_j^2 + 1 \\
        &= (\beta'\delta'_a + \delta_y')^2 + (\beta'\gamma'_a + \gamma_y')^2 + {\beta'}^2\var(\epsilon_a') + \var(\epsilon_y'),  
    \end{align}
    and plugging in  \eqref{matching:UY} yields
    \begin{align}
        \sum\beta_j^2 + 1 
        &= {\beta'}^2\var(\epsilon_a') + \var(\epsilon_y'), 
    \end{align}
    which simplifies to: 
    \begin{align}
        \var(&\epsilon_y') = \sum\beta_j^2 + 1 - {\beta'}^2\var(\epsilon_a')  
        = \sum\beta_j^2 + 1 - \frac{(\sum_j \alpha_j\beta_j)^2}{\sum_j \alpha_j^2} \\
        &=1 + \frac{\sum_j\beta_j^2\sum_j\alpha_j^2 - (\sum_j \alpha_j\beta_j)^2}{\sum_j \alpha_j^2} 
        = 1 + \frac{\sum_{l<j}(\beta_l\alpha_j - \beta_j\alpha_l)^2}{\sum_j \alpha_j^2}, 
    \end{align}
    where the last equality is by Lagrange's identity. 
    In summary, the aggregate IV SCM~\eqref{scm:generalcase} is equivalent to SCM~\eqref{scm:classic_erv_IV} where
    \begin{align}
        \beta' &= \frac{\sum_j \alpha_j\beta_j}{\sum_j \alpha_j^2}, \quad 
        \delta_a' = \sum_j \alpha_j\delta_j, \quad  
        \gamma_a' = \sum_j \alpha_j\gamma_{a_j}, \\
        \delta_y' &= \sum_j \beta_j\delta_j -  \frac{(\sum_j \alpha_j\beta_j)( \sum_j \alpha_j\delta_j)}{\sum_j \alpha_j^2}, \\ 
        \gamma_y' &= \sum_j \beta_j\gamma_{a_j} + \gamma_y -  \frac{(\sum_j \alpha_j\beta_j)(\sum_j \alpha_j\gamma_{a_j})}{\sum_j \alpha_j^2},
     \end{align}
    \begin{align}
        \var(\epsilon_a') &= \sum_j \alpha_j^2, \quad \var(\epsilon_y') = 1 + \frac{\sum_{l<j}(\beta_l\alpha_j - \beta_j\alpha_l)^2}{\sum_j \alpha_j^2}.
    \end{align}

    In the proportional aggregation case ($\beta_j = \tau\alpha_j$, $j\in\{1, \dots, k\}$), SCM~\eqref{scm:classic_erv_IV} collapses to the classic IV model: 
    \begin{align}
    \beta' &= \frac{\sum_j \alpha_j\beta_j}{\sum_j \alpha_j^2} = \frac{\tau(\sum\alpha_j^2)}{\sum\alpha_j^2} = {\tau},\\
       \delta_y' &= \sum_j \beta_j\delta_j - \beta'(\sum_j \alpha_j\delta_j) = \sum_j \beta_j\delta_j - {\tau}\left(\frac{1}{\tau}\bigg[\sum_j \beta_j\delta_j\bigg]\right) = 0.
    \end{align}
\end{proof}

\section{Calculations Regarding the Sargan Test}\label{app:sargan-fig}

\subsection{Calculations for Table~\ref{table:instrument_strength}}

Under SCM~\eqref{eq:sargan0}, $ \var(A) = \left(\delta_{11} + \delta_{12}\right)^2 + \left(\delta_{21} + \delta_{22}\right)^2  + 3$, and $\cov(I_l, A) = \delta_{l1} + \delta_{l2},  \enspace l = 1,2$.
Thus the last two columns of Table~\ref{table:instrument_strength} are calculated using
\begin{align}
    \cor(I_l, A) = \frac{\cov(I_l, A)}{\text{sd}(I_l)\text{sd}(A)} = \frac{\delta_{l1} + \delta_{l2}}{\sqrt{\left(\delta_{11} + \delta_{12}\right)^2 + \left(\delta_{21} + \delta_{22}\right)^2  + 3}}, \quad l=1,2. 
\end{align}

\begin{figure}
\centering
\includegraphics[width = 12cm]{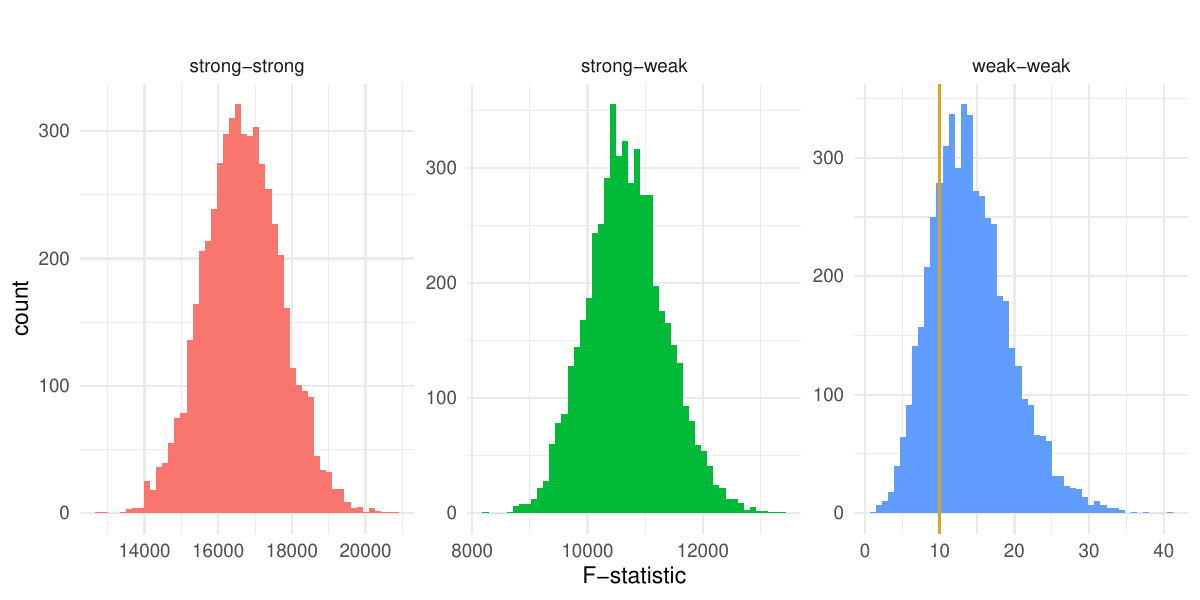}
\caption{Distribution of $F$-statistics for pairs of instruments of different strengths. The typical cut-off of 10 is marked in yellow where applicable.}
\label{fig:F_stats_distn}
\end{figure}

\section{Extensions: Aggregating Other Variables}\label{appendix:extensions}

\begin{figure}[H]
\centering
\includegraphics[width=9cm]{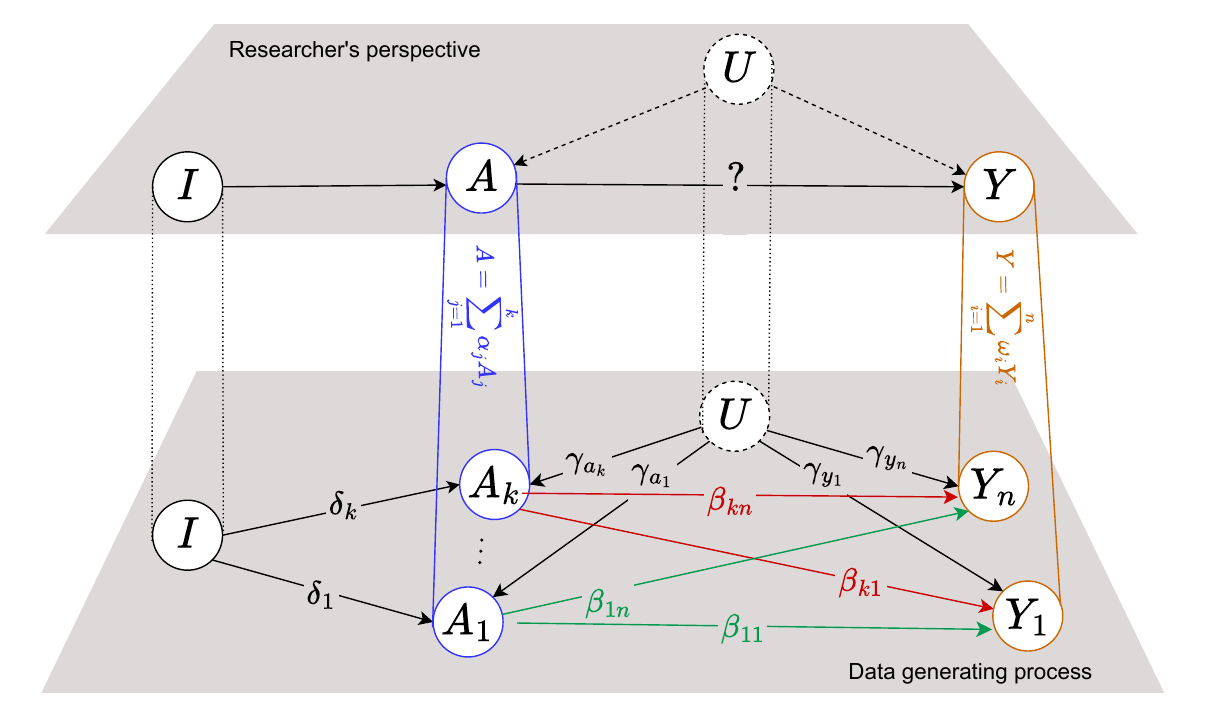}
\caption{Aggregate setting with aggregate outcome $Y$ and treatment $A$.}
         \label{fig:case1_yaggr}
\end{figure} 

\subsection{Aggregate Treatment and Aggregate Outcome}\label{sec:case1_yaggr}
We now consider an extension to SCM~\eqref{scm:generalcase}, visualized in the bottom of Figure~\ref{fig:case1_yaggr}:
\begin{align}
    Y = \sum_{i=1}^m \omega_i Y_i, \quad 
    Y_i  \leftarrow \sum_{j=1}^k \beta_{ji} A_j + \gamma_{yi}U + \epsilon_{yi} \label{scm:case1_yaggr}
\end{align}
We have not added any new information to $A$ nor $I$ and the first stage calculation for $\tilde{A}$ remains the same. The second stage is altered by the fact that
\begin{align}
    \cov(Y,\tilde A) \enspace &= \-\ \cov\left(\sum_{i=1}^m \sum_{j=1}^k \omega_{i}(\beta_{ji} A_j + \gamma_{yi}U + \epsilon_{yi}), \frac{\cov(\sum_j \alpha_jA_j,I)}{\var(I)} I\right) \\
&= \frac{\cov(\sum_j \alpha_jA_j,I)}{\var(I)} \sum_{i=1}^m \sum_{j=1}^k \omega_{i}\beta_{ji}\delta_j\var(I). 
\end{align}
This alteration produces a new IV estimand, $\beta^y_{\text{IV}}$: 
\begin{align}
\betaIV^y &= \sum_{i=1}^m \omega_{i} \frac{\sum_{j} \beta_{ji}\delta_j}{\sum_j \alpha_j\delta_j}. \label{eq:2slsyaggr}
\end{align}
$\beta^y_{\text{IV}}$ is a weighted sum of IV estimands of the form  $\betaIV$, where each estimand and weight corresponds to an outcome component $Y_i$ and its related discussions in the main text.

\subsection{Aggregate Treatment and Aggregate Instrument}

 \begin{figure}[H]
\centering
\includegraphics[width = 9cm]{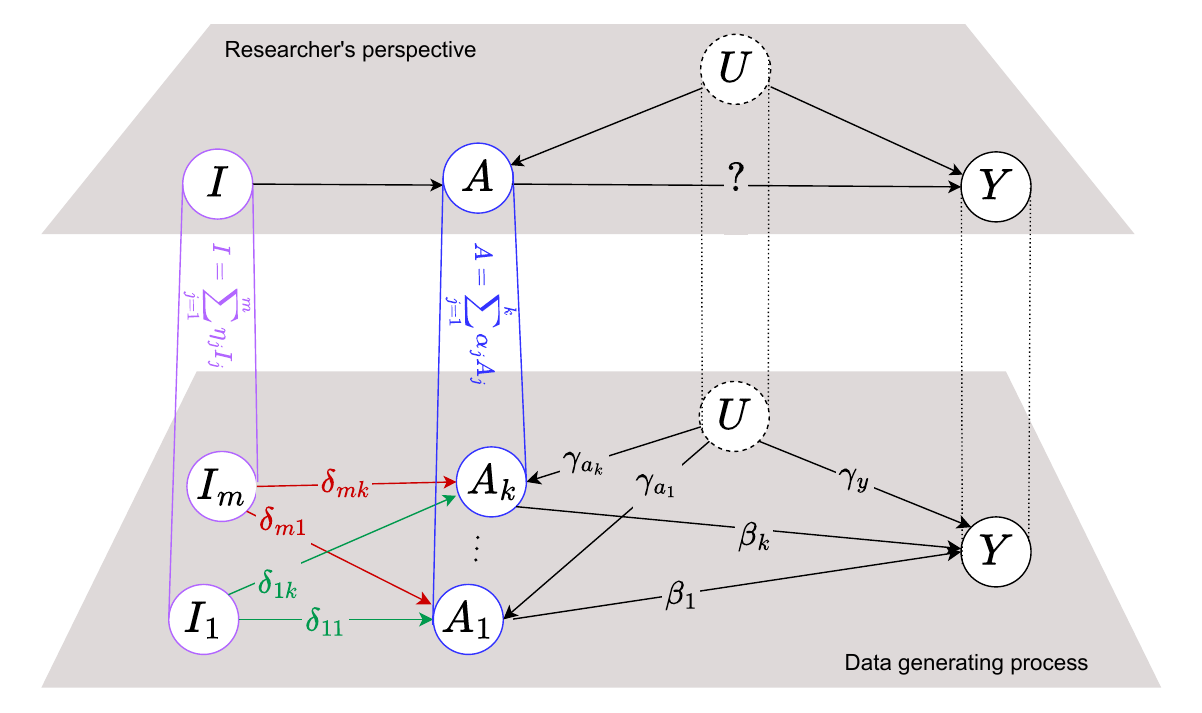}
\caption{Aggregate setting with aggregate instrument $I$ and treatment $A$.}
        \label{fig:case1_iaggr}
\end{figure}
We now consider an aggregate instrument $I$. With reference to Section~\ref{sec:case1_yaggr}, we consider $Y$ as a non-aggregate variable without loss of generality. We refer to  Figure~\ref{fig:case1_iaggr} and the structural equations below for details.
\begin{align}
    I &= \sum_{j=1}^m \eta_j I_j \label{scm:case1_iaggr_I} \\
    A_i  &\leftarrow \sum_{j=1}^k \delta_{ji} I_j + \gamma_{a_i}U + \epsilon_{a_i} \label{scm:case1_iaggr}
\end{align}
Unlike the aggregate $Y$ case, we are unable to decompose the 2SLS calculation into more granular calculations (i.e. one for each of $I_1,...,I_k$). Instead, consider the OLS projection of $A_i$ onto $I$: 
\begin{align}\label{eq:OLS_AI}
A_i = \xi_i I + \epsilon_{a_i}
\end{align}
such that $\xi_i = \frac{\cov(A_i,I)}{\var(I)}$ and $\cov(\epsilon_{a_i}, I) = 0$. Then $\betaIV^I$ is just a modified version of $\betaIV$ from \eqref{eq:2sls} where $\delta_1, ..., \delta_k$ are replaced with the OLS coefficients from \eqref{eq:OLS_AI}: 
\begin{align}
    \beta^I_{\text{IV}} &= 
    \frac{\cov(Y,I)}{\cov(A,I)}  
    = \frac{\cov\left(\sum_{j=1}^k \beta_j A_j ,I\right)}{\cov\left(\sum_{j=1}^k \alpha_j A_j,I\right)}  
    = \frac{\sum_{j=1}^k \beta_j\xi_j}{\sum_{j=1}^k \alpha_j\xi_j}.
\end{align}
Thus the discussions in the main text once more apply, with the alteration that the IV estimand is no longer a function of a single instrument's effects on the $A_i$ but is rather a function of the OLS estimands from regressing each $A_i$ on the aggregate $I$. 

We note that IV estimation provides a valid estimand when $A$ is not an aggregate but $I$ and $Y$ are aggregates. Indeed, performing IV estimation with an aggregate instrument $I$ and components $I_1, ..., I_k$ (and non-aggregate $A$) means that instead of regressing each of $A$ and $Y$ on the instrument vector $(I_1, ..., I_k)^\top$, we have chosen to regress on the fixed linear combination 
$
I = \sum_{j=1}^k \eta_j I_j.
$
As IV estimation only requires the instrument $I$ to be associated with $A$, substituting such an $I$  raises no issues. An aggregated outcome $Y$ (and non-aggregate $A$)  implies a straightforward interpretation of the IV estimand as a weighted sum of the individual effects of the treatment on the outcome components, where the weights are the components' contributions to the aggregate.

\end{document}